\documentclass[twocolumn]{revtex4}
\usepackage[dvips]{graphicx}
\usepackage{float, color,array,amsfonts}

\begin{document}

\title{Spin and thermal transport and critical phenomena in three-dimensional antiferromagnets}

\author{Kazushi Aoyama}

\date{\today}

\affiliation{Department of Earth and Space Science, Graduate School of Science, Osaka University, Osaka 560-0043, Japan}

\begin{abstract}
We investigate spin and thermal transport near the N\'{e}el transition temperature $T_N$ in three dimensions, by numerically analyzing the classical antiferromagnetic $XXZ$ model on the cubic lattice, where in the model, the anisotropy of the exchange interaction $\Delta=J_z/J_x$ plays a role to control the universality class of the transition. It is found by means of the hybrid Monte-Carlo and spin-dynamics simulations that in the $XY$ and Heisenberg cases of $\Delta \leq 1$, the longitudinal spin conductivity $\sigma^s_{\mu\mu}$ exhibits a divergent enhancement on cooling toward $T_N$, while not in the Ising case of $\Delta>1$. In all the three cases, the temperature dependence of the thermal conductivity $\kappa_{\mu\mu}$ is featureless at $T_N$, being consistent with experimental results. The divergent enhancement of $\sigma^s_{\mu\mu}$ toward $T_N$ is attributed to the spin-current relaxation time which gets longer toward $T_N$, showing a power-law divergence characteristic of critical phenomena. It is also found that in contrast to the $XY$ case where the divergence in $\sigma^s_{\mu\mu}$ is rapidly suppressed below $T_N$, $\sigma^s_{\mu\mu}$ likely remains divergent even below $T_N$ in the Heisenberg case, which might experimentally be observed in the ideally isotropic antiferromagnet RbMnF$_3$.   
\end{abstract}

\maketitle
\section{Introduction}
In magnetic materials, dynamical properties of interacting spins, such as magnetic excitations and fluctuations, are often reflected in transport phenomena where the properties of the electric and thermal currents have widely been discussed. Recently, thanks to the development of experimental methods in the context of spintronics \cite{Spincurrent-mag_Frangou_16, Spincurrent-mag_Qiu_16, Spincurrent-mag_Frangou_17, Spincurrent-mag_Gladii_18, Spincurrent-mag_Ou_18}, the spin current is also becoming available as a probe to study the spin dynamics. This demands us to explore the fundamental physics underlying the association between the spin transport and magnetic phase transitions. Previously, we theoretically investigated transport properties of two-dimensional insulating magnets and showed that the $XY$-type magnetic anisotropy leads to a divergence in the longitudinal spin conductivity $\sigma^s_{\mu\mu}$ at the Kosterlitz-Thouless (KT) transition temperature \cite{trans-sq_AK_prb_19}.  
In this paper, we extend our analysis to a three-dimensional system and numerically investigate the spin and thermal transport near the antiferromagnetic transition whose critical behavior is controlled by a magnetic anisotropy.

As is well known, magnetic phase transitions can be described by classical spin models, and a magnetic anisotropy plays a role to control the universality class of the transition. A minimal model possessing the magnetic anisotropy would be the classical nearest-neighbor (NN) $XXZ$ model which is given by 
\begin{equation}\label{eq:Hamiltonian}
{\cal H} = - J\sum_{\langle i,j \rangle} \Big( S^x_i S^x_j + S^y_i S^y_j + \Delta S^z_i S^z_j  \Big), 
\end{equation}
where $S^\alpha_i$ is $\alpha$-component of a classical spin ${\bf S}_i$ at a lattice site $i$, $\langle i,j\rangle$ denotes the summation over all the NN pairs, $J<0$ is the NN antiferromagnetic exchange interaction, and $\Delta > 0$ is a dimensionless parameter characterizing the magnetic anisotropy. For simplicity, we consider unfrustrated systems where the ground state is the two-sublattice N\'{e}el order. 
In the case of the two-dimensional square lattice, a second-order antiferromagnetic transition and the KT transition \cite{KT_KT_73} occur at finite temperatures for the Ising-type ($\Delta >1$) and $XY$-type ($\Delta <1$) anisotropies, respectively, while in the isotropic Heisenberg case of $\Delta=1$, a phase transition does not occur at any finite temperature \cite{Heisenberg_Polyakov_75}. In the case of the three-dimensional cubic lattice, the Ising-type, $XY$-type, and Heisenberg-type spin systems commonly undergo a second-order antiferromagnetic transition at a finite temperature $T_N$, but their critical properties such as the exponents of the power-law behaviors in various physical quantities depend on $\Delta$ \cite{3Dall_Pelissetto_pr_02}. Since as exemplified by the critical slowing down, the spin dynamics is generally affected by the phase transition \cite{3Dall_dynamical_rmp_77}, characteristic transport phenomena may appear near $T_N$.  

\begin{figure*}[t]
\includegraphics[scale=0.75]{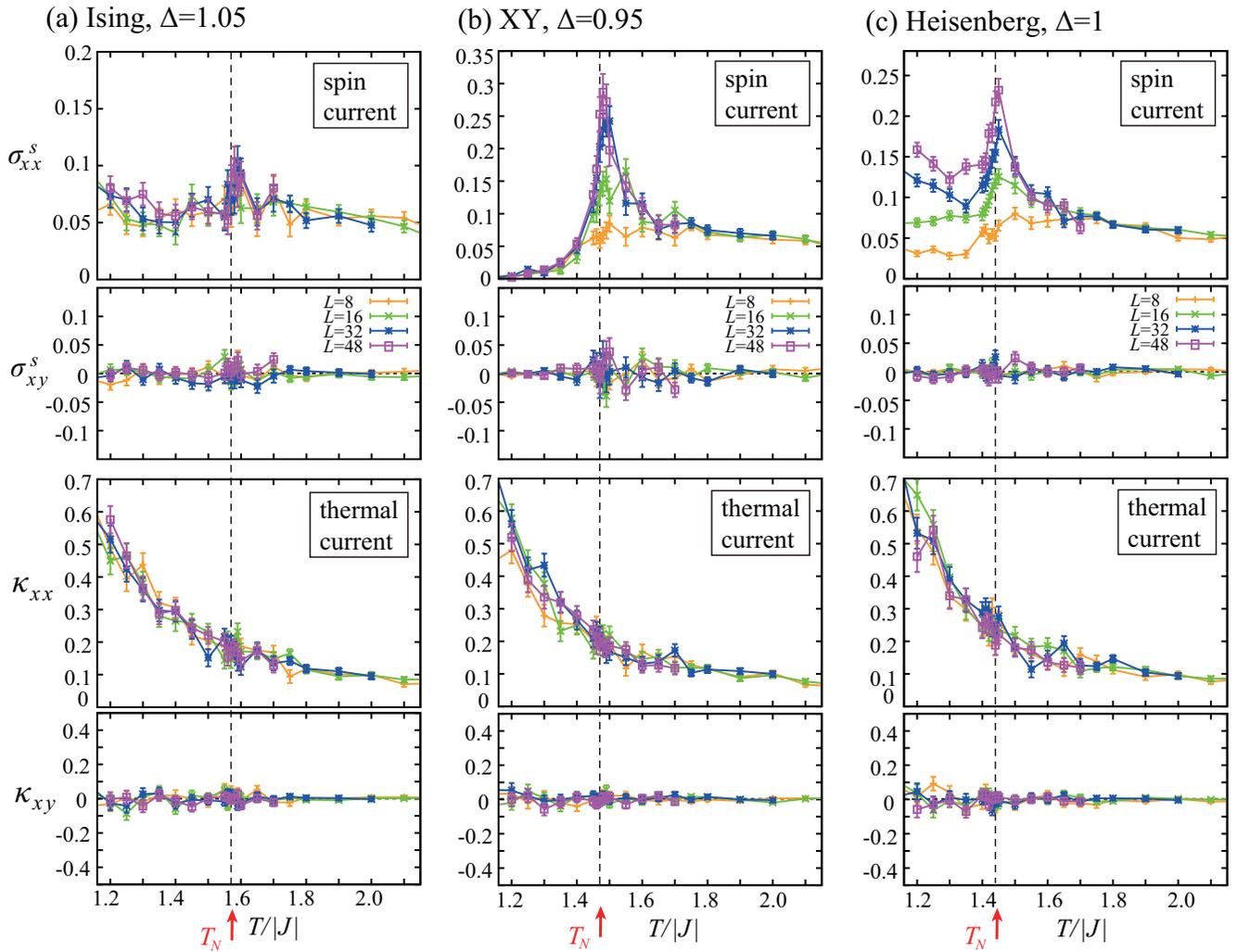}
\caption{Temperature dependence of the spin conductivity $\sigma^s_{\mu \nu}$ (the first and second panels from the top, corresponding to the longitudinal and transverse conductivities, respectively) and the thermal conductivity $\kappa_{\mu \nu}$ (the third and fourth panels from the top, corresponding to the longitudinal and transverse conductivities, respectively) near the antiferromagnetic transition temperature $T_N$ in the (a) Ising-type ($\Delta=1.05$), (b) $XY$-type ($\Delta=0.95$), and (c) Heisenberg-type ($\Delta=1$) spin systems. $\sigma^s_{\mu \nu}$ is a dimensionless quantity and $\kappa_{\mu \nu}$ is measured in units of $|J|$. In (a), (b) and (c), dashed lines indicate $T_N/|J|\simeq 1.574$, 1.472, and 1.443, respectively (see Fig. \ref{fig:basic_comp} in Appendix A). \label{fig:trans_comp}}
\end{figure*}

In our previous paper, we numerically demonstrated that in two dimensions, the difference in the ordering properties is reflected in the spin-current transport, while not for the thermal current. In the $XY$ case, the longitudinal spin conductivity $\sigma^s_{\mu\mu}$ exhibits a divergent enhancement toward the KT topological transition associated with binding-unbinding of magnetic vortices, whereas in the Ising and Heisenberg cases, it only shows almost monotonic temperature dependences \cite{trans-sq_AK_prb_19}. Our result, i.e., the enhancement of $\sigma^s_{\mu\mu}$ at the KT transition temperature, is supported by a later analytical approach \cite{KTtrans_Kim_21}, and a similar divergent enhancement can also be found in the frustrated triangular-lattice Heisenberg antiferromagnet \cite{trans-tri_AK_prl_20} where a KT-like binding-unbinding transition of the $\mathbb{Z}_2$ vortices is expected to occur \cite{Z2_Kawamura_84, Z2_Kawamura_10, Z2_Kawamura_11, Z2_Tomiyasu_22}. Then, the question is how the critical phenomena in three dimensions are reflected in the transport properties. In this paper, we investigate temperature dependences of the conductivities of the spin and thermal currents in the Ising-type ($\Delta > 1$), $XY$-type ($\Delta < 1$), and Heisenberg-type ($\Delta = 1$) antiferromagnets on the cubic lattice by means of the hybrid Monte-Carlo (MC) and spin-dynamics simulations.

Our result near the antiferromagnetic transition temperature $T_N$ is summarized in Fig. \ref{fig:trans_comp}, where the upper (lower) two panels show the temperature dependence of the spin conductivity $\sigma^s_{\mu\nu}$ (the thermal conductivity $\kappa_{\mu\nu}$). As readily seen from the top panels in Fig. \ref{fig:trans_comp}, in the $XY$ and Heisenberg cases of $\Delta \leq 1$, the longitudinal spin conductivity $\sigma^s_{xx}$ ($=\sigma^s_{yy}=\sigma^s_{zz}$) shows a divergent enhancement on cooling toward $T_N$, whereas in the Ising case of $\Delta>1$, it only shows a slight enhancement. Furthermore, although in the $XY$ case, the divergence in $\sigma^s_{xx}$ is rapidly suppressed below $T_N$, it remains divergent even below $T_N$ in the Heisenberg case where $\sigma^s_{xx}$ increases with increasing the system size $L$, suggesting $\sigma^s_{xx}\rightarrow \infty$ in the thermodynamic limit of $L\rightarrow \infty$. In contrast to such characteristic behaviors in the spin transport, the longitudinal thermal conductivity $\kappa_{xx}$ ($=\kappa_{yy}=\kappa_{zz}$) increases monotonically without showing a divergent anomaly in all the three cases of $\Delta>1$, $\Delta<1$, and $\Delta=1$ (see the third panels from the top in Fig. \ref{fig:trans_comp}), as is actually the case for experimental results on relevant magnets \cite{FeF2_Marinelli_prb_95, RbMnF3_Marinelli_prb_96}.
The Hall responses $\sigma^s_{xy}$ and $\kappa_{xy}$ are absent over the whole temperature range (see the second and fourth panels from the top in Fig. \ref{fig:trans_comp}). The significant enhancement of $\sigma^s_{xx}$ toward $T_N$ turns out to be associated with the spin-current relaxation time $\tau_s$ which gets longer toward $T_N$, showing a power-law divergence characteristic of the critical phenomena.

This paper is organized as follows: In Sec. II, the theoretical framework to calculate the transport coefficients in magnetic insulators will be explained. Numerical results on the spin and thermal transports will be discussed in detail in Secs. III and IV, respectively, where the properties not only near $T_N$ but also below $T_N$ will be addressed. We end this paper with summary and discussion in Sec. V. For reference, MC results on the fundamental static physical quantities in the present $XXZ$ model and the analytical results on the low-temperature transport properties in the linear spin-wave theory (LSWT) are shown in Appendixes A and B, respectively.    

\section{Theoretical framework}
Since the expressions of the spin and thermal currents in the $XXZ$ model and the formulas to calculate their conductivities in the linear response theory have already been derived elsewhere \cite{trans-sq_AK_prb_19}, here, we will briefly summarize the procedure how to calculate the spin and thermal conductivities $\sigma^s_{\mu \nu}$ and $\kappa_{\mu\nu}$. It should be emphasized here that the spin dynamics equation and the current expressions can be derived directly from the spin Hamiltonian (\ref{eq:Hamiltonian}) and thereby, no assumption has been made except the spin Hamiltonian.

\subsection{Spin dynamics}
For the Hamiltonian (\ref{eq:Hamiltonian}), the spin dynamics, i.e., the time evolution of the spins, is determined by the following equation of motion: 
\begin{eqnarray}\label{eq:Bloch}
\frac{d {\bf S}_i}{dt} &=&  {\bf S}_i \times {\bf H}_i^{\rm eff}, \nonumber\\
{\bf H}_i^{\rm eff} &=& J\sum_{j \in N(i)} \big( S^x_j, S^y_j, \Delta S^z_j  \big), 
\end{eqnarray}
where $N(i)$ denotes all the NN sites of $i$. Since Eq. (\ref{eq:Bloch}) is a classical analogue of the Heisenberg equation for the spin operator, all the static and dynamical magnetic properties purely intrinsic to the Hamiltonian (\ref{eq:Hamiltonian}) should be described by the combined use of Eqs. (\ref{eq:Hamiltonian}) and (\ref{eq:Bloch}). Equation (\ref{eq:Bloch}) corresponds to the Landau-Lifshitz-Gilbert (LLG) equation \cite{LLG_Landau_35} without a phenomenological damping term. Note that as our starting point is the spin Hamiltonian (\ref{eq:Hamiltonian}) without couplings to other degrees of freedom such as phonons and conduction electrons, the extrinsic damping term does not appear in Eq. (\ref{eq:Bloch}). Thus, the spin and current relaxations are due to thermal fluctuations whose nature is determined by the Hamiltonian (\ref{eq:Hamiltonian}). 

\begin{figure}[t]
\includegraphics[width=\columnwidth]{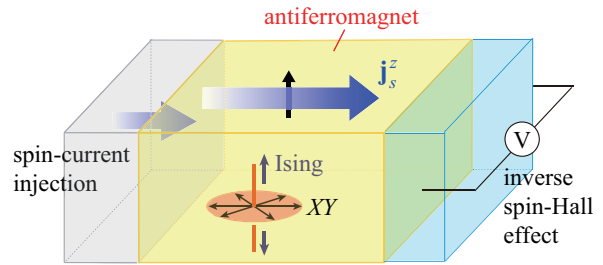}
\caption{Possible experimental setup for measuring the longitudinal spin conductivity $\sigma^s_{\mu\mu}$ in a bulk antiferromagnet (yellow region), where the spin current ${\bf j}_s^z$ (blue arrow) has its spin polarization parallel to the uniaxial direction (easy and hard axes in the Ising and $XY$ cases, respectively). The spin current is injected from a ferromagnet or a metal (gray region) by using a spin pumping or the spin-Hall effect, respectively, and is detected as an electric signal in a metal such as Pt (sky-blue region) by using the inverse spin-Hall effect \cite{injection-detection_review_Hou_19}. \label{fig:setup}}
\end{figure}
\subsection{Conductivities of spin and thermal currents}
In general, a conserved physical quantity of the system ${\cal O}=\int d{\bf r} \,  {\cal O}({\bf r},t)$ should satisfy the continuity equation $\frac{\partial}{\partial t}{\cal O}({\bf r},t)+\nabla \cdot {\bf j}_{\cal O}({\bf r}, t)=0$ with associated local current density ${\bf j}_{\cal O}({\bf r}, t)$, so that one has 
\begin{equation}
\int d{\bf r} \, {\bf r} \frac{\partial}{\partial t}{\cal O}({\bf r},t) = -\int d{\bf r} \, {\bf r} \,\nabla \cdot {\bf j}_{\cal O}({\bf r}, t) = \int d{\bf r} \, {\bf j}_{\cal O}({\bf r}, t).
\end{equation} 
Thus, the net current ${\bf J}_{\cal O}(t)$ is given by \cite{book_Mahan}
\begin{equation}
 {\bf J}_{\cal O}(t)=\int d{\bf r} \, {\bf j}_{\cal O}({\bf r}, t) = \int d{\bf r} \, {\bf r} \frac{\partial}{\partial t}{\cal O}({\bf r},t).
\end{equation}
In the present $XXZ$ model, the $z$ component of the magnetization $M^z=\sum_i S^z_i$ and the total energy ${\cal H} = \sum_i {\cal H}_i$ with ${\cal H}_i=\frac{-J}{2}\sum_{j \in N(i)}\big( S^x_i S^x_j + S^y_i S^y_j + \Delta S^z_i S^z_j  \big)$ are conserved, so that the associated currents, namely, the spin and thermal currents (${\bf J}^z_s$ and ${\bf J}_{th}$) are given by \cite{trans-sq_AK_prb_19}
\begin{equation}\label{eq:j_sc}
{\bf J}^z_s(t) = \sum_i {\bf r}_i \frac{d S^z_i}{d t} = J\sum_{\langle i,j \rangle} \big({\bf r}_i-{\bf r}_j \big) \big( {\bf S}_i \times {\bf S}_j \big)^z, 
\end{equation}
\begin{eqnarray}\label{eq:j_th}
{\bf J}_{th}(t) &=& \sum_i {\bf r}_i \frac{-J}{2}\sum_{j \in N(i)}\frac{d}{dt}\big( S^x_iS^x_j+S^y_i S^y_j+\Delta S^z_i S^z_j \big) \nonumber\\
&=& \frac{J^2}{4}\sum_i  \sum_{j,k \in N(i)} \big( {\bf r}_j - {\bf r}_k \big) \Big\{({\bf S}_j\times{\bf S}_k)^z S^z_i \nonumber\\
&& + \Delta \Big[ ({\bf S}_j \times {\bf S}_k)^x S^x_i + ({\bf S}_j \times {\bf S}_k)^y S^y_i \Big] \Big\},
\end{eqnarray} 
where Eq. (\ref{eq:Bloch}) has been used in replacing the time derivative $\frac{dS^\alpha_i}{dt}$ with a product $S^\beta_j \, S^\gamma_k$. It turns out that ${\bf J}^z_s$ and ${\bf J}_{th}$ are related to the vector spin chirality ${\bf S}_i \times {\bf S}_j$ and the scalar spin chirality ${\bf S}_i\cdot ({\bf S}_j\times {\bf S}_k)$, respectively \cite{SpinDyn_Huber_74,SpinDyn_Jencic_prb_15,MHall_Mook_prb_16,MHall_Mook_prb_17,Thermal_Huber_ptp_68,SpinDyn_Zotos_prb_05,SpinDyn_Kawasaki_68,SpinDyn_Sentef_07,SpinDyn_Pires_09,SpinDyn_Chen_13}. We note that in the presence of the magnetic anisotropy, only the uniaxial $z$ component of the magnetization is conserved, so that the associated spin current ${\bf j}^z_s$ has its polarization along the uniaxial direction, i.e., easy and hard axes in the Ising and $XY$ cases, respectively (see Fig. \ref{fig:setup}). 

In general, the spin and thermal currents are obtained as responses of the magnetic-field and temperature gradients, respectively \cite{MHall_Mook_prb_16, MHall_Mook_prb_17, trans-sq_AK_prb_19}. In real spin-current measurements, however, as shown in Fig. \ref{fig:setup}, the spin current may be injected into the bulk antiferromagnet from a ferromagnet or a metal by using a spin pumping or the spin-Hall effect, respectively \cite{injection-detection_review_Hou_19}. The spin conductivity $\sigma^s_{\mu\mu}$ could be measured by detecting the transmitted spin current in the opposite side as an electric signal via the inverse spin-Hall effect.
Within the linear response theory \cite{KuboFormular_Kubo_57}, the spin and thermal conductivities in bulk magnets are given by 
\begin{eqnarray}\label{eq:conductivity}
\sigma_{\mu \nu}^s &=& \frac{1}{T \, L^3} \int_0^\infty dt \, \big\langle J^z_{s,\nu}(0) \, J^z_{s,\mu}(t) \big\rangle, \\
\kappa_{\mu \nu} &=& \frac{1}{T^2 \, L^3} \int_0^\infty dt \, \big\langle J_{th,\nu}(0) \, J_{th,\mu}(t) \big\rangle, \nonumber
\end{eqnarray} 
where $L$ is a linear system size and $\langle {\cal O} \rangle$ denotes the thermal average of a physical quantity ${\cal O}$.  
In the Heisenberg case of $\Delta=1$ where the spin space is isotropic, not only the $z$ component of the magnetization but also the $x$ and $y$ components are conserved, so that one can also define the spin currents ${\bf J}_s^x$ and ${\bf J}_s^y$ as well as ${\bf J}_s^z$ all of which are equivalent to one another because of the isotropic nature of the spin space. Thus, in the Heisenberg case, we calculate the spin conductivity averaged over the three spin components
\begin{equation}\label{eq:conductivity_Heisenberg}
\sigma_{\mu \nu}^s  =  \frac{1}{T \, L^3} \frac{1}{3} \sum_{\alpha=x,y,z} \int_0^\infty dt \,  \big\langle J^\alpha_{s,\nu}(0) \, J^\alpha_{s,\mu}(t) \big\rangle 
\end{equation}
instead of Eq. (\ref{eq:conductivity}). 

Now, the problem is reduced to calculate the time correlations of the spin and thermal currents $\langle J^z_{s,\nu}(0) \, J^z_{s,\mu}(t) \rangle$ and $\langle J_{th,\nu}(0) \, J_{th,\mu}(t) \rangle$ at various temperatures. For the present cubic lattice, the total number of spin $N_{\rm spin}$ and the system size $L$ are related by $L^3=N_{\rm spin} \, a^3$ with lattice constant $a$. As the time $t$ is measured in units of $|J|^{-1}$, it turns out that $\sigma^s_{\mu \nu}$ and $\kappa_{\mu \nu}$ have the dimension of $1/a$ and $|J|/a$, respectively. Throughout this paper, we take $a=1$ for simplicity. 

\subsection{Numerical method}
The time evolutions of ${\bf J}^z_s$ and ${\bf J}_{th}$ are determined microscopically by the spin-dynamics equation (\ref{eq:Bloch}). By numerically integrating Eq. (\ref{eq:Bloch}), we calculate the time correlations $\langle J^z_{s,\nu}(0) \, J^z_{s,\mu}(t) \rangle$ and $\langle J_{th,\nu}(0) \, J_{th,\mu}(t) \rangle$ at each time step. In the numerical integration of Eq. (\ref{eq:Bloch}), we use the second order symplectic method which guarantees the exact energy conservation \cite{Symplectic_Krech_98}. We have partly checked that the results obtained here are not altered if the 4th order Runge-Kutta method is used instead. To properly evaluate the integral over time in Eq. (\ref{eq:conductivity}), we perform long-time integrations typically up to $t=10\,|J|^{-1}$ at high temperatures above $T_N$ and $600 \, |J|^{-1}$ at the lowest temperature with the time step $\delta t=0.01 \, |J|^{-1}$ until the time correlations $\big\langle J^z_{s,\nu}(0) \, J^z_{s,\mu}(t) \big\rangle$ and $\big\langle J_{th,\nu}(0) \, J_{th,\mu}(t) \big\rangle$ are completely lost.

To incorporate temperature effects, we use temperature-dependent equilibrium spin configurations as the initial states for the equation of motion (\ref{eq:Bloch}), and the thermal average is taken as the average over initial equilibrium spin configurations generated in the MC simulations.    
In this work, at each temperature $T$, we prepare 8000 equilibrium spin configurations by picking up a spin snapshot in every 100 MC sweeps after 10$^5$ MC sweeps for thermalization in 8 independent runs, where our one MC sweep consists of the 1 heat-bath sweep and successive 10 over-relaxation sweeps. 

By analyzing the system-size dependences of the spin conductivity $\sigma^s_{\mu \nu}$ and the thermal conductivity $\kappa_{\mu \nu}$ at given temperatures, we discuss the temperature dependences of $\sigma^s_{\mu \nu}$ and $\kappa_{\mu \nu}$ in the thermodynamic limit ($L \rightarrow \infty$) of our interest. In the present cubic lattice where $x$, $y$, and $z$ directions are equivalent to one another, the relations $\sigma^s_{xx}=\sigma^s_{yy}=\sigma^s_{zz}$ and $\kappa_{xx}=\kappa_{yy}=\kappa_{zz}$ trivially hold, and such a situation is also the case for the transverse conductivities $\sigma^s_{\mu \nu}$ and $\kappa_{\mu \nu}$ with $\mu\neq \nu$. Thus, in this work, we only discuss the $xx$ and $xy$ components, $\sigma^s_{xx}$, $\sigma^s_{xy}$, $\kappa_{xx}$, and $\kappa_{xy}$. 

In this work, the magnetic anisotropy $\Delta$ is only one system parameter: $\Delta > 1$, $\Delta < 1$, and $\Delta=1$ correspond to the Ising-type, $XY$-type, and Heisenberg-type spin systems, respectively. Throughout this paper, the parameter values of $\Delta=1.05$ and $\Delta=0.95$ are used for the Ising and $XY$ cases, respectively, as typical values slightly deviating from $\Delta=1$ for the isotropic Heisenberg case. From the MC simulations (see Appendix A), $T_N$ in each case can be estimated as $T_N/|J| \simeq 1.574$ for $\Delta=1.05$, $T_N/|J| \simeq 1.472 $ for $\Delta=0.95$, and $T_N/|J| \simeq 1.443$ for $\Delta=1$.

\section{Spin conductivity $\sigma^s_{\mu\nu}$}
\begin{figure}[t]
\includegraphics[width=\columnwidth]{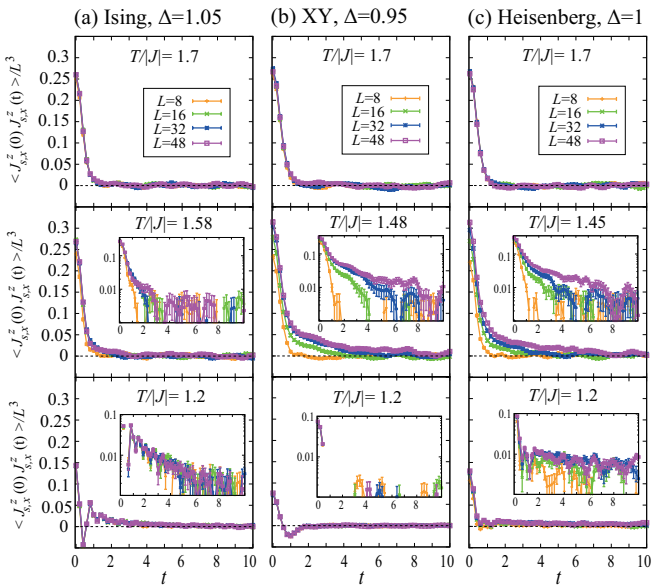}
\caption{The time correlation function of the spin current $\langle J^x_{s,x}(0) \, J^z_{s,x}(t) \rangle/L^3$ at $T=1.7|J|>T_N$ (top), $T \sim T_N$ (middle), and $T=1.2|J|<T_N$ (bottom) in the (a) Ising-type ($\Delta=1.05$ and $T_N/|J| \simeq 1.574$), (b) $XY$-type ($\Delta=0.95$ and $T_N/|J| \simeq1.472$), and (c) Heisenberg-type ($\Delta=1$ and $T_N/|J| \simeq 1.443$) spin systems. In the middle panels, $T/|J|$'s in (a), (b), and (c) are $1.58$, $1.48$, and $1.45$, respectively. Time $t$ and $\langle J^z_{s,x}(0) \, J^z_{s,x}(t) \rangle/L^3$ are measured in units of $|J|^{-1}$ and $|J|^2$, respectively. In the middle and bottom panels, the inset shows a semilogarithmic plot of the main panel.  \label{fig:timedep_spin_comp}}
\end{figure}

In this section, we will discuss the association between the spin transport and the antiferromagnetic transition, based on numerical results obtained in the Ising-type ($\Delta=1.05$), $XY$-type ($\Delta = 0.95$), and Heisenberg-type ($\Delta=1$) spin systems. The main focus is on how the differences in the universality class and the critical magnetic fluctuation are reflected in the spin conductivity $\sigma^s_{\mu\nu}$.

As mentioned in Sec. I, the Hall response corresponding to the transverse conductivity $\sigma^s_{xy}$ is absent over the whole temperature range, we will focus on the longitudinal spin conductivity $\sigma^s_{xx}$ as a representative example of the three equivalent $\sigma^s_{xx}$, $\sigma^s_{yy}$, and $\sigma^s_{zz}$. We will show that the longitudinal spin conductivity $\sigma^s_{xx}$ exhibits a divergent enhancement toward $T_N$ in the $XY$ and Heisenberg cases, while not in the Ising case. Although our main interest is in the spin transport near $T_N$, for completeness, we will also discuss its low-temperature behavior below $T_N$ where the spin waves or magnons should carry the current. 

As the fundamental information of the temperature dependence of $\sigma^s_{\mu\nu}$ consists in the time correlation of the spin current $\langle J^z_{s,\nu}(0) \, J^z_{s,\mu}(t) \rangle$ except the trivial $T^{-1}$ factor [see Eq. (\ref{eq:conductivity})], we will start from the temperature dependence of $\langle J^z_{s,x}(0) \, J^z_{s,x}(t) \rangle$. 

\subsection{Time correlation function} 
Figure \ref{fig:timedep_spin_comp} shows the time correlation function normalized by the system size $\langle J^z_{s,x}(0) \, J^z_{s,x}(t) \rangle/L^3$ at $T>T_N$ (top panels), $T\sim T_N$ (middle panels), and $T<T_N$ (bottom panels) in the (a) Ising-type ($\Delta = 1.05$), (b) $XY$-type ($\Delta = 0.95$), and (c) Heisenberg-type ($\Delta=1$) spin systems. 
At the high temperature $T/|J|=1.7$ sufficiently above $T_N$, the time correlation rapidly decays in all the three-types of spin systems (see the top panels in Fig. \ref{fig:timedep_spin_comp}). With decreasing temperature, differences among the three cases become clearer. At a temperature close to but slightly above $T_N$, the relaxation time gets longer with increasing the system size $L$ in the $XY$ and Heisenberg cases, whereas in the Ising case, it is saturated for larger sizes (see the middle panels in Fig. \ref{fig:timedep_spin_comp}). This suggests that in the thermodynamic limit of $L \rightarrow \infty$, the relaxation time is very long in the $XY$ and Heisenberg cases, while not in the Ising case. As one can see from the bottom panels in Fig. \ref{fig:timedep_spin_comp}, at the low temperature $T/|J|=1.2$ below $T_N$, the time correlation function commonly shows an oscillating behavior or a dip structure in a short-time scale, and in a long-time scale, it slowly decays in the Ising and Heisenberg cases, whereas in the $XY$ case, the time correlation is completely lost (see the semi-logarithmic plots shown in the insets). The slowly decaying long-time tail is system-size dependent in the Heisenberg case, while not in the Ising case. As will be explained below, these features of the spin-current relaxation are reflected in the temperature dependence of the spin conductivity $\sigma^s_{xx}$.

\subsection{Longitudinal spin conductivity $\sigma^s_{\mu\mu}$ near $T_N$}
We will first discuss overall qualitative  features of $\sigma^s_{\mu\nu}$ near $T_N$. As shown in the upper two panels in Fig. \ref{fig:trans_comp}, although the transverse Hall response $\sigma^s_{xy}$ is absent in all the three-types of spin systems, the longitudinal spin conductivity $\sigma^s_{xx}$ exhibits characteristic temperature dependences depending on the value of the magnetic anisotropy $\Delta$. In the Ising case of $\Delta=1.05$, $\sigma^s_{xx}$'s for larger $L$'s are almost system-size independent as expected  from the almost $L$-independent time-correlation-function in Fig. \ref{fig:timedep_spin_comp} (a), so that they correspond to the thermodynamic-limit ($L\rightarrow\infty$) value which only shows a slight enhancement near $T_N$. In the $XY$ case of $\Delta=0.95$, $\sigma^s_{xx}$ exhibits a divergent sharp peak toward $T_N$, and becomes vanishingly small below $T_N$. Since the peak height increases with increasing the system size $L$, $\sigma^s_{xx}$ should diverge at $T_N$ in the thermodynamic limit. In the Heisenberg case of $\Delta=1$, $\sigma^s_{xx}$ also exhibits a similar divergent behavior toward $T_N$, but even below $T_N$, it remains system-size dependent and increases with increasing $L$, suggesting that in the thermodynamic limit, $\sigma^s_{xx}$ may be infinite over the low-temperature region below $T_N$. 

\begin{figure}[t]
\includegraphics[width=\columnwidth]{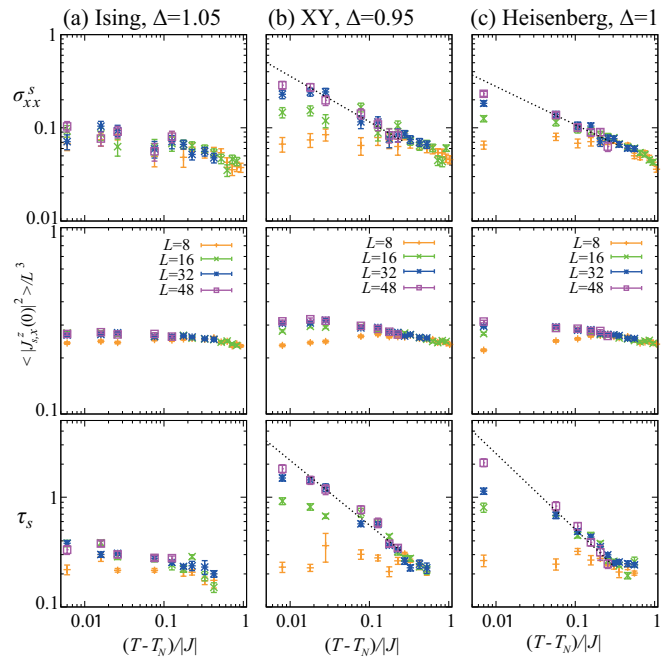}
\caption {The log-log plot of the longitudinal spin conductivity $\sigma^s_{x x}$ (top), the equal-time correlation function of the spin current $\langle |J^z_{s,x}(0)|^2 \rangle/L^3$ (middle), and the spin-current relaxation time $\tau_s$ (bottom) as a function of $(T-T_{N})/|J|$ in the (a) Ising-type, (b) $XY$-type, and (c) Heisenberg-type spin systems. A dotted line represents a power function of $(T-T_N)$ obtained by fitting the size-independent data in the temperature range of $T_N<T<1.7 \, |J|$ (for details, see the text). \label{fig:Js_nearTN}}
\end{figure}

Below in this subsection, we will focus on the increasing behavior of  $\sigma^s_{xx}$ toward $T_N$ on cooling from above. The top panels in Fig. \ref{fig:Js_nearTN} show the log-log plot of $\sigma^s_{xx}$ as a function of $(T-T_N)/|J|$ in the (a) Ising, (b) $XY$, and (c) Heisenberg cases. The regular plot of Fig. \ref{fig:Js_nearTN} in a wider temperature range is shown in Fig. \ref{fig:Js_allT} which will be discussed in the next subsection. In the log-log plot of $\sigma^s_{xx}$ in Fig. \ref{fig:Js_nearTN}, the larger-size data for the $XY$ and Heisenberg cases increase toward $T_N$ almost linearly as indicated by dotted lines, suggestive of the power-law divergence of the form $c_\sigma \, (T-T_N)^{-x_\sigma}$. By fitting the size-independent data with this functional form, we obtain the exponent $x_\sigma$ as $x_\sigma=0.49$ and 0.41 for the $XY$ and Heisenberg cases, respectively. In the Ising case, on the other hand, the $\sigma^s_{xx}$ value is almost saturated on approaching $T_N$, so that $\sigma^s_{xx}$ is non-divergent. 

Noting that the $S^x$ and $S^y$ spin components play an essential role for the spin current ${\bf J}_s^z$ in the form of $({\bf S}_i \times {\bf S}_j)^z$ [see Eq. (\ref{eq:j_sc})], we could understand the origin of the above difference between the Ising and other two cases as follows: in the $XY$ and Heisenberg cases, the spin fluctuations in the $S^xS^y$ plane perpendicular to the polarization of the spin current (see Fig. \ref{fig:setup}) become critical, leading to the significant enhancement of $\sigma^s_{xx}$, while not in the Ising case where only the longitudinal mode along the $S^z$ direction, which should be irrelevant to the spin current  ${\bf J}_s^z$, develops.  The slight enhancement of $\sigma^s_{xx}$ near $T_N$ in the Ising case of $\Delta=1.05$ would be due to the remnant Heisenberg nature which is gradually smeared out on approaching $T_N$, showing a crossover to the Ising universality class. Thus, it should be suppressed for larger values of the magnetic anisotropy $\Delta$, as is actually the case for the associated two-dimensional system (see Fig. 8 in Ref. \cite{trans-sq_AK_prb_19})

Now, we shall move on to the origin of the power-law divergence of $\sigma^s_{xx}$ in the $XY$ and Heisenberg cases. Since $\sigma^s_{xx}$ is obtained by integrating the time correlation function $\langle J^z_{s,x}(0) \, J^z_{s,x}(t) \rangle/L^3$ over time, $\langle |J^z_{s,\nu}(0)|^2 \rangle/L^3$ as well as the spin-current relaxation time $\tau_s$ should be important. In Fig. \ref{fig:timedep_spin_comp}, the time correlation $\langle J^z_{s,\nu}(0) \, J^z_{s,\mu}(t) \rangle/L^3$ decays exponentially in the form of $e^{- t/\tau_s}$, so that we could assume $\langle J^z_{s,x}(0) \, J^z_{s,x}(t) \rangle/L^3 \simeq \big( \langle |J^z_{s,x}(0)|^2 \rangle/L^3 \big) \, e^{-t/\tau_{s}}$. Then, by carrying out the integral over time in Eq. (\ref{eq:conductivity}), one can estimate the longitudinal spin conductivity as $\sigma^s_{xx} \simeq T^{-1} \, \tau_s \, \langle |J^z_{s,x}(0)|^2 \rangle/L^3 $. Bearing this relation in our mind, we will discuss the origin of the power-law divergence.

The middle and bottom panels in Fig. \ref{fig:Js_nearTN} show the log-log plot of $\langle |J^z_{s,x}(0)|^2 \rangle/L^3$ and $\tau_s$, respectively, as a function of $(T-T_N)/|J|$. One can see that $\tau_s$ and $\sigma^s_{xx}$ show similar temperature and system-size dependences (compare top and bottom panels in Fig. \ref{fig:Js_nearTN}) and that the temperature dependence of $\langle |J^z_{s,x}(0)|^2 \rangle/L^3$ is relatively weak. In the $XY$ and Heisenberg cases, $\tau_s$ shows a power-law divergence similarly to $\sigma^s_{xx}$, as indicated by dotted lines in the bottom panels in Fig. \ref{fig:Js_nearTN}, where the dotted lines are obtained by fitting the size-independent $\tau_s$ data in the temperature range of $T_N<T<1.7|J|$ with a power-law function $c_\tau \, (T-T_N)^{-x_\tau}$. The obtained exponent of $x_\tau=0.59$ (0.69) in the $XY$-type (Heisenberg-type) spin system is relatively close to the exponent for the spin conductivity $x_\sigma=0.49$ (0.41), suggesting that the power-law divergence of the spin conductivity $\sigma^s_{xx}$ is attributed to the spin-current relaxation time $\tau_s$ which gets longer toward $T_N$ to eventually diverge. In each spin system, a slight deviation between the two exponents would be due to the non-divergent temperature dependence of $\langle |J^z_{s,x}(0)|^2 \rangle/L^3$ and the trivial $T^{-1}$ factor appearing in $\sigma^s_{xx}$. If one can evaluate $\sigma^s_{xx}$ and $\tau_s$ in the temperature region further close to $T_N$ where the critical divergence becomes much clearer, further close values of the exponents could be obtained, but it needs further larger-size simulations.
 
As the power-law divergences of $\sigma^s_{xx}$ and $\tau_s$ are numerically confirmed, next question is how their exponents are related to the critical exponents associated with the three-dimensional N\'{e}el transition. In the three-dimensional Ising, $XY$, and Heisenberg universality classes, the critical exponents $\nu$'s characterizing the divergence of the spin correlation length $\xi_s \sim (T-T_N)^{-\nu}$ are known to be $\nu = 0.630$, 0.671, and 0.711, respectively \cite{3Dall_Pelissetto_pr_02, 3DHeisenberg_Campostrini_02, 3DXY_Campostrini_01}. The dynamical critical exponent $z$ characterizing the divergence of the spin correlation time $\tau \sim \xi_s^z \sim (T-T_N)^{-z \, \nu}$ generally depends on the sign of $J$, and is given by $z \simeq 2.18$ ($z=3/2$) for the antiferromagnetic Ising (Heisenberg) system \cite{3DHeisenberg_dynamical_Kawasaki_68, 3DHeisenberg_dynamical_Halperin_69, 3DHeisenberg_dynamical_Tsai_03}, which corresponds to the $z$ value for Model C (G) in Ref. \cite{3Dall_dynamical_rmp_77}. In the $XY$ case, $z=1.5$ is expected for ferromagnetic $J>0$ \cite{3Dall_dynamical_rmp_77, 3DXY_Thoma_prb_91, 3DXY_Krech_prb_99}, but the corresponding value for antiferromagnetic $J<0$ is not available. Thus, for the moment, we {\it assume} that the value of $z=1.5$ is also satisfied for $J<0$.
Then, the net exponent $z\nu$ for the time scale of the critical slowing down is calculated as $z\nu\simeq 1.0$ and $1.06$ in the $XY$ and Heisenberg cases, respectively. The $z\nu$ values are not so far from the associated exponents for the spin-current relaxation time, 0.59 and 0.69, and the spin conductivity, 0.49 and 0.41, but we cannot rule out the possibility that the time scales of the spin itself and the spin current may be different. Actually, it is indicated that in the $XY$ and Heisenberg cases, the critical behaviors in the spin conductivity are roughly described by $\xi_s^{2-z}\sim (T-T_N)^{-0.34}$ \cite{3DXY_Krech_prb_99} and $\xi_s^{1/2} \sim (T-T_N)^{-0.36}$ \cite{3DHeisenberg_dynamical_Kawasaki_68}, respectively, whose exponents and our result also do not differ so much.  
Although it is difficult to provide a quantitative argument on the critical exponent for the spin conductivity $\sigma^s_{xx}$ or the spin-current relaxation time $\tau_s$, it is certain that $\sigma^s_{xx}$ and $\tau_s$ diverge toward $T_N$ due to the transverse spin fluctuation associated with the critical phenomena. 

Although $\sigma^s_{xx}$'s in both the $XY$-type and Heisenberg-type spin systems exhibit the divergence at $T_N$, their low-temperature properties below $T_N$ are quite different. As will be explained below, in the former anisotropic case, $\sigma^s_{xx}$ is rapidly suppressed to zero, whereas in the latter isotropic case, $\sigma^s_{xx}$ likely remains divergent over the wide temperature range below $T_N$.

\subsection{Longitudinal spin conductivity $\sigma^s_{\mu\mu}$ below $T_N$}
\begin{figure}[t]
\includegraphics[width=\columnwidth]{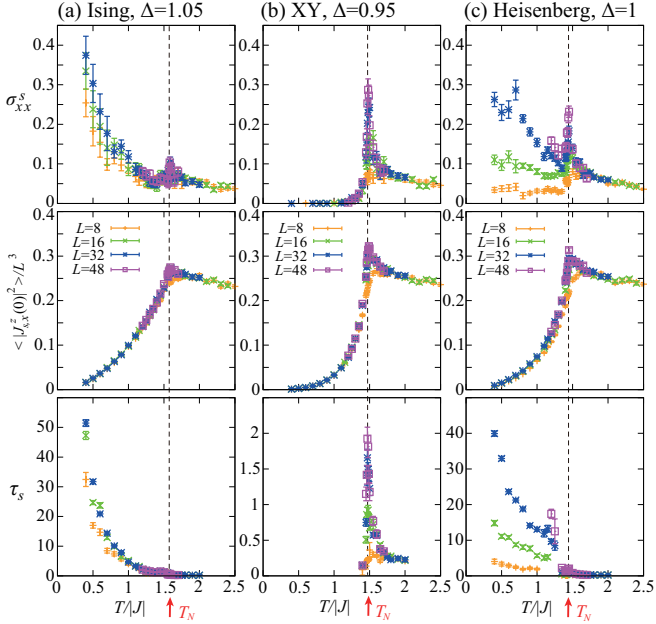}
\caption {The temperature dependence of the longitudinal spin conductivity $\sigma^s_{x x}$ (top), the equal-time correlation function of the spin current $\langle |J^z_{s,x}(0)|^2 \rangle/L^3$ (middle), and the spin-current relaxation time $\tau_s$ (bottom) in the (a) Ising-type, (b) $XY$-type, and (c) Heisenberg-type spin systems. In (a), (b) and (c), dashed lines indicate $T_N/|J|\simeq 1.574$, 1.472, and 1.443, respectively. Top panels are the same as those in Fig. \ref{fig:trans_comp} except the temperature range; in this figure, the temperature range is extended to a sufficiently low temperature below $T_N$. \label{fig:Js_allT}}
\end{figure}

Below $T_N$ where the long-range antiferromagnetic order is developed, the spin-waves or magnons should be relevant to the spin and thermal transport. 
In the present classical spin model, quantum effects, which in real materials, govern the low-temperature magnon excitation in the form of the Bose distribution function, are inherently absent. In this respect, the transport properties of the classical spin systems in the $T\rightarrow 0$ limit should be unrealistic. On the other hand, at moderate temperatures below $T_N$ where the quantum effect is masked by the thermal fluctuation, the classical description may work well. In this subsection, bearing this temperature range in our mind, we will discuss the spin transport below $T_N$.  
Before discussing the numerical result, we will summarize the analytical result obtained in the linear spin-wave theory for the present classical spin system (for details, see Appendix B). 

First, the magnon excitation is gapless in the $XY$ and Heisenberg cases of $\Delta \leq 1$, while not in the Ising case of $\Delta >1$ where the easy-axis magnetic anisotropy yields the excitation gap [see Eq. (\ref{eq:magenergy_app}) in Appendix B 1]. It turns out that in the Ising and Heisenberg cases of $\Delta \geq 1$, the equal-time correlation of the magnon-spin-current $\langle |J^z_{s,x}(0)|^2 \rangle/L^3$ gradually decreases with decreasing temperature, showing a $T^2$ dependence [see Eq. (\ref{eq:Js_static_tmp}) in Appendix B 2], whereas in the $XY$ case of $\Delta <1$, it vanishes because the leading-order magnon-spin-current is absent [see Eq. (\ref{eq:current_spin_mag}) in Appendix B 1]. Concerning the longitudinal spin conductivity $\sigma^s_{xx}$ mediated by the magnons [see Eq. (\ref{eq:conductivity_classical_spin}) in Appendix B 3], it is roughly proportional to $T/\alpha$ in the Ising case, where $\alpha$ denotes the magnon damping of its origin consisting in the spin Hamiltonian (\ref{eq:Hamiltonian}), and is known to show a $T^2$ dependence at least in the classical isotropic case \cite{MagnonDamping_Harris_71}. Thus, for a weak Ising anisotropy, $\sigma^s_{xx}\propto T^{-1}$ is expected. In the $XY$ case, $\sigma^s_{xx}$ is zero because the leading-order magnon-spin-current is absent from the beginning. In the Heisenberg case, $\sigma^s_{xx}$ involves a logarithmic divergence, so that $\sigma^s_{xx}$ is infinite over the low-temperature ordered phase where the magnons are well-defined.

Now, we will discuss the numerical result for $T<T_N$.
Figure \ref{fig:Js_allT} shows the temperature dependence of $\sigma^s_{xx}$, $\langle |J^z_{s,x}(0)|^2 \rangle/L^3$, and $\tau_s$ over the wide temperature range including a sufficiently low temperature below $T_N$. Note that zoomed views of the top panels near $T_N$ correspond to the top panels in Fig. \ref{fig:trans_comp}. In the bottom panels in Figs. \ref{fig:Js_allT} (b) and (c), there are no data points in the wide and narrow temperature regions just below $T_N$, respectively. In the former case, the time correlation decays too fast, so that we cannot evaluate such a very short $\tau_s$ within our precision, whereas in the latter case, the decay function does not look like a simple exponential form and thus, $\tau_s$ cannot uniquely be determined in this narrow temperature region.

In the Ising case shown in Fig. \ref{fig:Js_allT} (a), on cooling across $T_N$, the spin conductivity $\sigma^s_{xx}$ is first suppressed just below $T_N$ and then, starts increasing toward $T=0$. Below $T_N$, since $\langle |J^z_{s,x}(0)|^2 \rangle/L^3$ is a decreasing function of $T$, the increasing behavior of $\sigma^s_{xx}$ is due to $\tau_s$ which gets longer toward $T=0$ [see the bottom panel of Fig. \ref{fig:Js_allT} (a)]. As the relation $\sigma^s_{xx} \simeq T^{-1} \, \tau_s \, \langle |J^z_{s,x}(0)|^2 \rangle/L^3 $ holds, these numerical results are qualitatively consistent with the analytical results for a {\it weak} Ising anisotropy,  $\langle |J^z_{s,x}(0)|^2 \rangle/L^3 \propto T^2$, $\sigma^s_{xx}\propto T^{-1}$, and $\tau_s \propto T^{-2}$. A similar lower-temperature behavior can also be seen in the associated two-dimensional system with the same anisotropy parameter of $\Delta=1.05$ \cite{trans-sq_AK_prb_19}. For larger $\Delta$, $\sigma^s_{xx}$ is gradually suppressed in the two-dimensional system as the excitation gap becomes larger \cite{trans-sq_AK_prb_19}. Such a situation would also be the case for the present three-dimensional system. 

In the $XY$ case shown in Fig. \ref{fig:Js_allT} (b), the spin conductivity $\sigma^s_{xx}$ diverges on cooling toward $T_N$, and once across $T_N$, it rapidly drops down to zero. Such a steep decrease can also be seen in $\langle |J^z_{s,x}(0)|^2 \rangle/L^3$ and $\tau_s$ [see the middle and bottom panels of Fig. \ref{fig:Js_allT} (b)], which is due to the fact that the leading-order magnon-spin-current is absent. In the associated two-dimensional $XY$-type spin system \cite{trans-sq_AK_prb_19}, the overall feature is quite similar to the present three-dimensional system except for the critical behavior above the transition where in two dimensions, the topological objects of vortices govern the physics \cite{KT_KT_73} . 

\begin{figure}[t]
\includegraphics[width=\columnwidth]{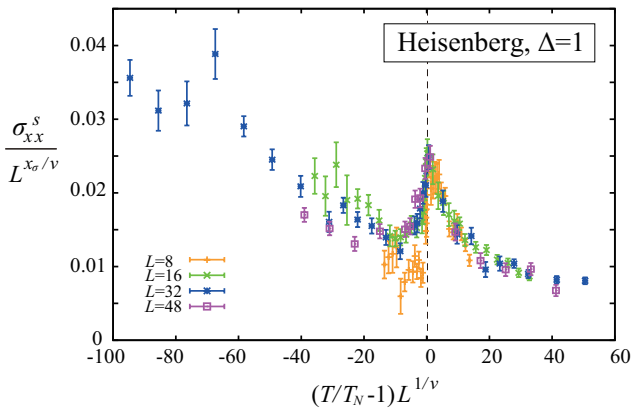}
\caption{The finite-size scaling plot of the longitudinal spin conductivity $\sigma^s_{xx}$ in the Heisenberg case of $\Delta=1$, where the well-established value of $\nu=0.711$ and the result of the present work $x_{\sigma}=0.41$ are used together with the transition temperature $T_N/|J|=1.443$. All the data in the top panel of Fig. \ref{fig:Js_allT} (c) are used in this scaling plot. \label{fig:scaling}}
\end{figure}

In the isotropic Heisenberg case shown in Fig. \ref{fig:Js_allT} (c), although $\langle |J^z_{s,x}(0)|^2 \rangle/L^3$ is system-size independent and gradually decreases toward $T=0$ (see the middle panel) as expected for the magnon spin current, both $\sigma^s_{xx}$ and $\tau_s$ are strongly size-dependent and their finite size data basically increase with decreasing temperature. 
Furthermore, at a fixed temperature below $T_N$, both $\sigma^s_{xx}$ and $\tau_s$ increases with increasing the system size $L$. If $\sigma^s_{xx}$ continues to increase even for larger $L$, this means that in the thermodynamic limit of $L\rightarrow \infty$, $\sigma^s_{xx}$ is divergent everywhere in the low-temperature ordered phase below $T_N$. Figure \ref{fig:scaling} shows the finite-size scaling plot of $\sigma^s_{xx}$ shown in the top panel of Fig. \ref{fig:Js_allT} (c). As readily seen, $\sigma^s_{xx}/L^{x_\sigma/\nu}$ can basically be scaled by a universal function $f\big((T/T_N-1)L^{1/\nu} \big)$, although the scaling is not so good for $T < T_N$. In the $L \rightarrow \infty$ limit, $f(x)$ tends to go to zero for $T_N < T$ ($x \rightarrow \infty$), while not for $T<T_N$ ($x\rightarrow - \infty$), suggesting that $\sigma^s_{xx} = L^{x_\sigma/\nu}f\big((T/T_N-1)L^{1/\nu}\big) \big|_{L\rightarrow \infty}$ is infinite for any $T<T_N$. The analytical calculation also supports this scenario. Thus, it is most likely that $\sigma^s_{xx}$ is infinite in the low-temperature long-range ordered phase with $\xi_s$ being infinite. We note that in the associated two-dimensional square-lattice system, $\sigma^s_{xx}$ is proportional to $\xi_s$ which exponentially increases toward $T=0$ but is finite at any finite temperature due to the dimensionality of the system \cite{trans-sq_AK_prb_19}. 
 
\section{Thermal conductivity $\kappa_{\mu\nu}$}
\begin{figure}[t]
\includegraphics[width=\columnwidth]{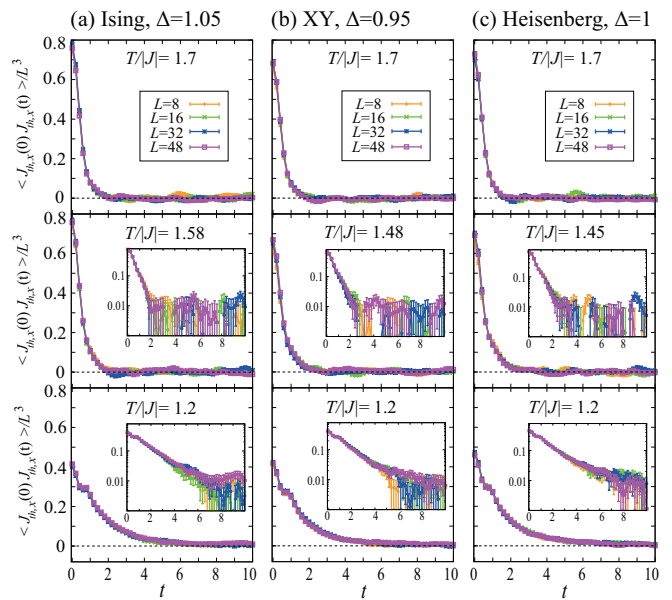}
\caption {The time correlation function of the thermal current $\langle J_{th,x}(0) \, J_{th,x}(t) \rangle/L^3$ for the same parameter sets as those in Fig. \ref{fig:timedep_spin_comp}. Time $t$ and $\langle J_{th,x}(0) \, J_{th,x}(t) \rangle/L^3$ are measured in units of $|J|^{-1}$ and $|J|^4$, respectively. In the middle and bottom panels, the inset shows a semilogarithmic plot of the main panel. \label{fig:timedep_thermal_comp}}
\end{figure}

In this section, we will discuss the thermal conductivity $\kappa_{\mu\nu}$. As readily seen from the bottom panels of Fig. \ref{fig:trans_comp}, the Hall response $\kappa_{xy}$ is absent over the whole temperature range, similarly to the transverse spin conductivity $\sigma^s_{xy}$, so that hereafter, we will focus on the longitudinal thermal conductivity $\kappa_{xx}$ ($=\kappa_{yy}=\kappa_{zz}$). We will show that in contrast to the spin conductivity $\sigma^s_{xx}$, the thermal conductivity $\kappa_{xx}$ only shows a monotonic increase on cooling across $T_N$. As $\kappa_{xx}$ is calculated from the time correlation function $\langle J_{th,x}(0) \, J_{th,x}(t) \rangle/L^3$ [see Eq. (\ref{eq:conductivity})], we shall start from the temperature dependence of $\langle J_{th,x}(0) \, J_{th,x}(t) \rangle/L^3$.

\subsection{Time correlation function}
Figure \ref{fig:timedep_thermal_comp} shows the time correlation function $\langle J_{th,x}(0) \, J_{th,x}(t) \rangle/L^3$ at various temperatures, where the same parameter sets as those in Fig. \ref{fig:timedep_spin_comp} have been used. There is no qualitative difference among the three cases, Ising, $XY$, and Heisenberg spin systems: the relaxation time of the thermal current gradually increases with decreasing temperature. Below $T_N$ (see the bottom panels of Fig. \ref{fig:timedep_thermal_comp}), $\langle J_{th,x}(0) \, J_{th,x}(t) \rangle/L^3$ shows a weak anomaly in the short-time scale, which might be related to the oscillating behavior in the spin-current relaxation shown in the bottom panels in Fig. \ref{fig:timedep_spin_comp}.  
Since near $T_N$, a long-time tail is commonly absent for the thermal-current relaxation (see the insets in Fig. \ref{fig:timedep_thermal_comp}), a critical anomaly is also absent in the associated thermal conductivity $\kappa_{xx}$, as will be explained below.

\subsection{Longitudinal thermal conductivity $\kappa_{\mu\mu}$ near $T_N$}
As shown in the third panels from the top in Fig. \ref{fig:trans_comp}, the longitudinal thermal conductivity $\kappa_{xx}$ monotonically increases on cooling with a slope steepening near $T_N$ in all the Ising, $XY$, and Heisenberg spin systems. Thus, in view of the main focus of this work, our conclusion is that the strong association between the thermal conductivity and the phase transition cannot be seen in three dimensions as well as in two dimensions \cite{trans-sq_AK_prb_19, trans-tri_AK_prl_20}. The present result of the non-divergent behavior of $\kappa_{xx}$ near $T_N$ is consistent with the experimental observation that in the antiferromagnets FeF$_2$ and RbMnF$_3$, which belong to the three-dimensional Ising and Heisenberg universality classes, respectively, $\kappa_{xx}$ only shows a non-divergent broad peak stemming from spin-phonon scatterings \cite{FeF2_Marinelli_prb_95, RbMnF3_Marinelli_prb_96}, which validates the present theoretical approach to calculate the transport coefficients in purely magnetic systems without coupling to other degrees of freedom such as phonons and electrons.

Since the thermal conductivity of {\it particles} is often expressed as $\kappa_{\mu\mu} \sim C \, v \, l$ with a particle velocity $v$ and a mean free path $l$, one may naively expect a characteristic behavior in $\kappa_{\mu\mu}$ similarly to the specific heat $C$. In the present case, however, the quasi-{\it particle} of the magnon is not well-defined for $T_N<T$ and the above expression cannot directly be applied in the temperature range across $T_N$, so that the temperature dependence of $\kappa_{\mu\mu}$ does not have to be the same as that of $C$. We note that this does not mean $\kappa_{\mu\mu}$ is always insensitive to a magnetic transition. Considering that in liquid $^4$He \cite{4He-thermal_Ahlers_prl_68}, $\kappa_{\mu\mu}$ diverges at the $\lambda$ transition belonging to the three-dimensional $XY$ universality class, the behavior of $\kappa_{\mu\mu}$ at the transition might depend on the sign of the exchange interaction, as in the case of the dynamical critical exponent. Our conclusion is that at least in the conventional antiferromagnetic insulators, there is no clear signature of the N\'{e}el transition in $\kappa_{\mu\mu}$.

In the low-temperature long-range ordered phase below $T_N$, the magnons should carry the thermal current, as in the case of the spin current. Since in the present classical spin system, the quantum effect in the form of the Bose distribution function is inherently absent, the low-temperature limit of the classical-spin thermal transport would not directly be related to realistic experimental situations. Nevertheless, to clarify the fundamental properties of the present classical system, we will discuss the low-temperature behavior of $\kappa_{xx}$ below $T_N$. 

\subsection{Longitudinal thermal conductivity $\kappa_{\mu\mu}$ below $T_N$}
As in the case of the spin conductivity $\sigma^s_{xx}$, the temperature dependence of the longitudinal thermal conductivity $\kappa_{xx}$ originates from that of the time correlation function of the thermal current $\langle J_{th,x}(0) \, J_{th,x}(t) \rangle/L^3$ except the trivial $T^{-2}$ factor [see Eq. (\ref{eq:conductivity})]. By using the equal-time correlation $\langle |J_{th,x}(0)|^2 \rangle/L^3$ and the relaxation time of the thermal current $\tau_{th}$ which can be deduced by fitting $\langle J_{th,x}(0) \, J_{th,x}(t) \rangle/L^3$ with the exponential form $e^{-t/\tau_{th}}$, one could write the thermal conductivity as $\kappa_{xx} \simeq T^{-2} \, \tau_{th} \, \langle |J_{th,x}(0)|^2 \rangle/L^3$. Below, we will discuss the $T$ dependence of $\kappa_{xx}$, $\langle |J_{th,x}(0)|^2 \rangle/L^3$, and $\tau_{th}$ toward $T=0$.

Before going to the numerical result, we will briefly summarize of the analytical result on the temperature dependence of the above quantities obtained in the linear spin-wave theory. For the magnon thermal current, $\langle |J_{th,x}(0)|^2 \rangle/L^3$ exhibits a $T^2$ dependence [see Eq. (\ref{eq:Jth_static}) in Appendix B 2], canceling the trivial $T^{-2}$ factor in $\kappa_{xx}$, so that $\kappa_{xx}$ is roughly proportional to the inverse of the magnon damping $\alpha$ [see Eq. (\ref{eq:conductivity_classical_th}) in Appendix B 3] and thereby, the thermal-current relaxation time $\tau_{th}$ is related to $\alpha$ via $\tau_{th} \propto 1/\alpha$. Since at least in the Heisenberg case, the magnon damping $\alpha$ is proportional to $T^2$, it follows that $\tau_{th} \propto T^{-2}$ and $\kappa_{xx} \propto T^{-2}$. Bearing these temperature dependences in our mind, we will discuss the numerical result.  

\begin{figure}[t]
\includegraphics[width=\columnwidth]{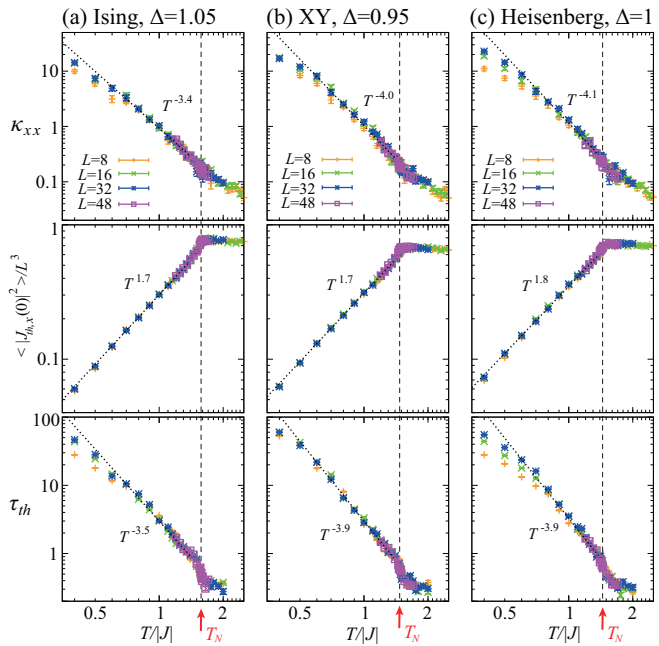}
\caption {The log-log plot of the temperature dependence of the longitudinal thermal conductivity $\kappa_{x x}$ (top), the equal-time correlation $\langle | J_{th,x}(0)|^2 \rangle/L^3$ (middle), and the relaxation time of the thermal current $\tau_{th}$ (bottom) in the (a) Ising-type ($\Delta=1.05$), (b) $XY$-type ($\Delta=0.95$), and (c) Heisenberg-type ($\Delta=1$) spin systems. A dashed line indicates $T_N$, and the temperature range is wider than that of Fig. \ref{fig:trans_comp}. In each panel, a dotted line represents a power-law function of the form $c \,T^{-x}$, where the exponent $x$ obtained by fitting the low-temperature data is explicitly written in the figure. \label{fig:thermal_Tdep}}
\end{figure}

Figure \ref{fig:thermal_Tdep} shows the log-log plot of $\kappa_{xx}$ (top), $\langle | J_{th,x}(0)|^2 \rangle/L^3$ (middle), and $\tau_{th}$ (bottom) in the Ising, $XY$, and Heisenberg cases. 
One can see from the top panels in Fig. \ref{fig:thermal_Tdep} that in the moderate temperature range below $T_N$, $0.7 \lesssim T/|J| \lesssim 1.4$, $\kappa_{xx}$ linearly increases toward $T=0$ in this log-log plot, suggesting a power-law behavior in this temperature range. By fitting the size-independent numerical data in this region with a power-law function $c \, T^{-x}$, we obtain the exponent $x$ as $x=3.4$, $4.0$, and $4.1$ in the Ising, $XY$, and Heisenberg cases, respectively, and the fitting results are indicated by dotted lines in Fig. \ref{fig:thermal_Tdep}. At further low temperatures, the increasing behavior of $\kappa_{xx}$ is slightly suppressed and the exponent $x$ tends to become smaller. Since $\langle | J_{th,x}(0)|^2 \rangle/L^3$ decreases roughly following the $T^2$ dependence expected for the magnon thermal-current (see the middle panels in Fig. \ref{fig:thermal_Tdep}), the increasing behavior of $\kappa_{xx}$ should originate from the thermal-current relaxation time $\tau_{th}$. Indeed, as one can see from the top and bottom panels in Fig. \ref{fig:thermal_Tdep}, $\kappa_{xx}$ and $\tau_{th}$ exhibit almost the same temperature dependence, and the exponents of the power-law behavior of $\tau_{th}$, which are obtained by the same fitting procedure as $x=3.5$, $3.9$, and $3.9$ in the Ising, $XY$, and Heisenberg cases, respectively, are almost the same as those for $\kappa_{xx}$.    

Compared with the analytically expected behavior of $\kappa_{xx} \propto \tau_{th} \propto 1/\alpha \propto T^{-2}$, the numerical result of $\kappa_{xx} \propto \tau_{th} \propto T^{-4}$ has a roughly twice larger exponent. This difference might be due to the temperature range we consider; the linear spin-wave theory is basically applicable to the lower-temperature region where the leading-order magnon contribution is important, whereas the fitting result of $T^{-4}$ is obtained in the moderate temperature range below $T_N$. The deviation from the $T^{-4}$ behavior at further low temperatures (see the top and bottom panels of Fig. \ref{fig:thermal_Tdep}) might be a signature of a crossover to the $T^{-2}$ dependence. In the associated two-dimensional Ising system, such a crossover behavior below $T_N$ can also be seen for the small value of $\Delta=1.05$, and for larger $\Delta$, the low-temperature $\kappa_{xx}$ is gradually suppressed due to the larger excitation gap \cite{trans-sq_AK_prb_19}. In the present three-dimensional system, the exponent of 3.5 for the Ising system is slightly smaller than the ones for the $XY$ and Heisenberg systems where the magnon excitation is gapless, which could be due to the gap opening.
    
\section{Summary and discussion}
We have theoretically investigated the spin and thermal transport near the N\'{e}el transition temperature $T_N$ in three-dimensional antiferromagnets by performing the hybrid Monte-Carlo and spin-dynamics simulations for the classical $XXZ$ model on the cubic lattice in which the anisotropy of the exchange interaction $\Delta \equiv J_z/J_x$ plays a role to control the universality class of the system. It is found that although the thermal conductivity $\kappa_{\mu\mu}$ is insensitive to the transition, being consistent with the experimental observations \cite{FeF2_Marinelli_prb_95, RbMnF3_Marinelli_prb_96}, the longitudinal spin conductivity $\sigma^s_{\mu\mu}$ is enhanced near $T_N$ with its temperature dependence being affected by the magnetic anisotropy $\Delta$: in the $XY$ ($\Delta <1$) and Heisenberg ($\Delta=1$) cases, $\sigma^s_{\mu\mu}$ diverges toward $T_N$ on cooling, while not in the Ising case ($\Delta>1$), suggesting that the magnetic fluctuation perpendicular to the polarization of the spin current is essential for the spin transport. The origin of the divergence in $\sigma^s_{\mu\mu}$ consists in the spin-current relaxation time $\tau_s$ which gets longer on approaching $T_N$ from above, and both $\sigma^s_{\mu\mu}$ and $\tau_s$ exhibit almost the same power-law divergences characteristic of critical phenomena. It is also found that in contrast to the $XY$ case where the divergence in $\sigma^s_{\mu\mu}$ is rapidly suppressed below $T_N$, $\sigma^s_{\mu\mu}$ likely remains divergent even below $T_N$ in the Heisenberg case of $\Delta=1$, pointing to the emergence of a ballistic/superdiffusion spin transport which has mainly been discussed in one-dimensional spin chains \cite{1Dsuperdiffusion_Ilievski_prl_18, 1Dsuperdiffusion_Gopalakrishnann_prl_19, 1Dsuperdiffusion_review_21}.

The above result for the three-dimensional system is qualitatively similar to that for the associated two-dimensional system, i.e., the classical antiferromagnetic $XXZ$ model on the square lattice \cite{trans-sq_AK_prb_19}. The common feature of the two systems is that in a situation where the transverse magnetic fluctuations are relevant to a phase transition, the longitudinal spin conductivity $\sigma^s_{\mu\mu}$ diverges at the transition temperature. This inversely suggests that the divergent enhancement of $\sigma^s_{\mu\mu}$ indicates a certain kind of a phase transition even if there is no clear anomaly in the static physical quantities such as the specific heat and magnetic susceptibility, as is actually the case for the KT transition in $XY$ antiferromagnets \cite{trans-sq_AK_prb_19} and the $\mathbb{Z}_2$-vortex transition in frustrated Heisenberg antiferromagnets \cite{trans-tri_AK_prl_20}. Thus, the spin current should serve as a probe of a transition in magnetic materials.    

Now, we address experimental implications of our result. In the spin-current measurements done on the antiferromagnets CoO and NiO in basically the same setting as that shown in Fig. \ref{fig:setup} \cite{Spincurrent-mag_Qiu_16}, the spin current injected from the Y$_3$Fe$_{5}$O$_{12}$ side by using the spin pumping is detected in the Pt side via the inverse spin-Hall effect, and the enhancement of the spin-current signal has been observed near $T_N$. It seems that CoO has an Ising-type easy-axis anisotropy \cite{CoO-NiO_aniso_Schron_prb_12, CoO-NiO_aniso_Roth_prl_58, CoO_aniso_Jauch_prb_01, CoO_aniso_Tomiyasu_jpsj_06}, whereas NiO has a biaxial anisotropy with a favorable direction in a $XY$-like easy-plane \cite{CoO-NiO_aniso_Schron_prb_12, NiO_aniso_Roth_pr_58, NiO_aniso_Kondoh_jpsj_64}. In the present theoretical work, we consider the situation where the spin polarization of the spin current is parallel to the uniaxial direction of the magnetic anisotropy (easy and hard axes in the Ising and $XY$ cases, respectively), because the magnetization is conserved for this polarization direction and thereby, the spin current is theoretically well-defined. Since the detailed information of the relative angle between the anisotropy axes of CoO and NiO and the polarization of the injected spin current is not available, at present, we cannot judge whether our result is consistent with the experimental observation or not. 
If one could perform a similar experiment on $XY$ and Heisenberg antiferromagnets such as SmMnO$_3$ \cite{SmMnO3_Oleaga_prb_12} and RbMnF$_3$ \cite{RbMnF3_Teaney_prl_62}, controlling the relative angle between the spin-current polarization and the anisotropy axis, the significant enhancement of the spin conductivity $\sigma^s_{\mu\mu}$ is expected to be observed at $T_N$. In particular, for the ideally isotropic antiferromagnet RbMnF$_3$ belonging to the three-dimensional Heisenberg universality class due to a very tiny magnetic anisotropy of the order of $10^{-6}$ \cite{RbMnF3_Teaney_prl_62, RbMnF3_Tucciarone_prb_71, RbMnF3_Kornblit_prb_73, RbMnF3_Coldea_prb_98, RbMnF3_Ropez_prb_14}, the high spin conductivity might persist even below $T_N$, as suggested from the present work.

Here, we comment on additional effects which are not incorporated in the present work but might be important in real experiments. 
First, in the setting shown in Fig. \ref{fig:setup}, effects of the interfaces between the antiferromagnet and both-side materials are not negligible. To capture the bulk signal undisturbed by the interface contribution, non-local measurements for thick antiferromagnets would be necessary \cite{afmIF_Takei_prb_15}. In addition, the efficiency of the spin-current injection and detection is determined by the spin-mixing conductance at the interfaces \cite{afmIF_Takei_prb_15, IF_Okamoto_prb_16, afmIF_Khymyn_prb_16, afmIF_Takei_prb_14}, being accompanied by a temperature dependence \cite{IF_Okamoto_prb_16}. Thus, a material combination having a relatively weak temperature dependence in the spin-mixing conductance would be better to see the change in the bulk $\sigma^s_{xx}$. Another factor which may possibly affect the conductivity measurement is the existence of phonons. In contrast to the thermal conductivity involving both magnetic and phonon contributions, however, the spin conductivity should be of purely magnetic origin unless a spin-phonon coupling is strong enough, with its high sensitivity to the critical phenomena associated with the magnetic transition. 

Although our focus in the present paper is on antiferromagnets, a divergent enhancement of the spin conductivity at a ferromagnetic transition is also indicated in the three-dimensional Heisenberg ferromagnet \cite{SpinDyn_Kawasaki_67}. In the ferromagnetic case, the dynamical critical exponents for the $XY$ and Heisenberg systems are known to be $z \simeq 1.5$ and $z\simeq 5/2$, respectively \cite{3DHeisenberg_dynamical_Tsai_03, 3DHeisenberg_dynamical_Halperin_69}. Such a large difference in $z$ may distinctly be reflected in the temperature dependence of the longitudinal spin conductivity $\sigma^s_{\mu\mu}$, shedding light on the association between the dynamical critical exponent $z$ and the exponent for the power-low divergence of $\sigma^s_{\mu\mu}$. We will leave this issue for our future work.

\begin{acknowledgments}
The author thanks Y. Niimi, H. Kawamura, and K. Tomiyasu for useful discussions. We are thankful to ISSP, the University of Tokyo and YITP, Kyoto University for providing us with CPU time. This work is supported by JSPS KAKENHI Grant Number JP21K03469.
\end{acknowledgments}

\appendix
\section{Ordering properties of the classical antiferromagnetic $XXZ$ model on the cubic lattice}
\begin{figure*}[t]
\includegraphics[scale=0.72]{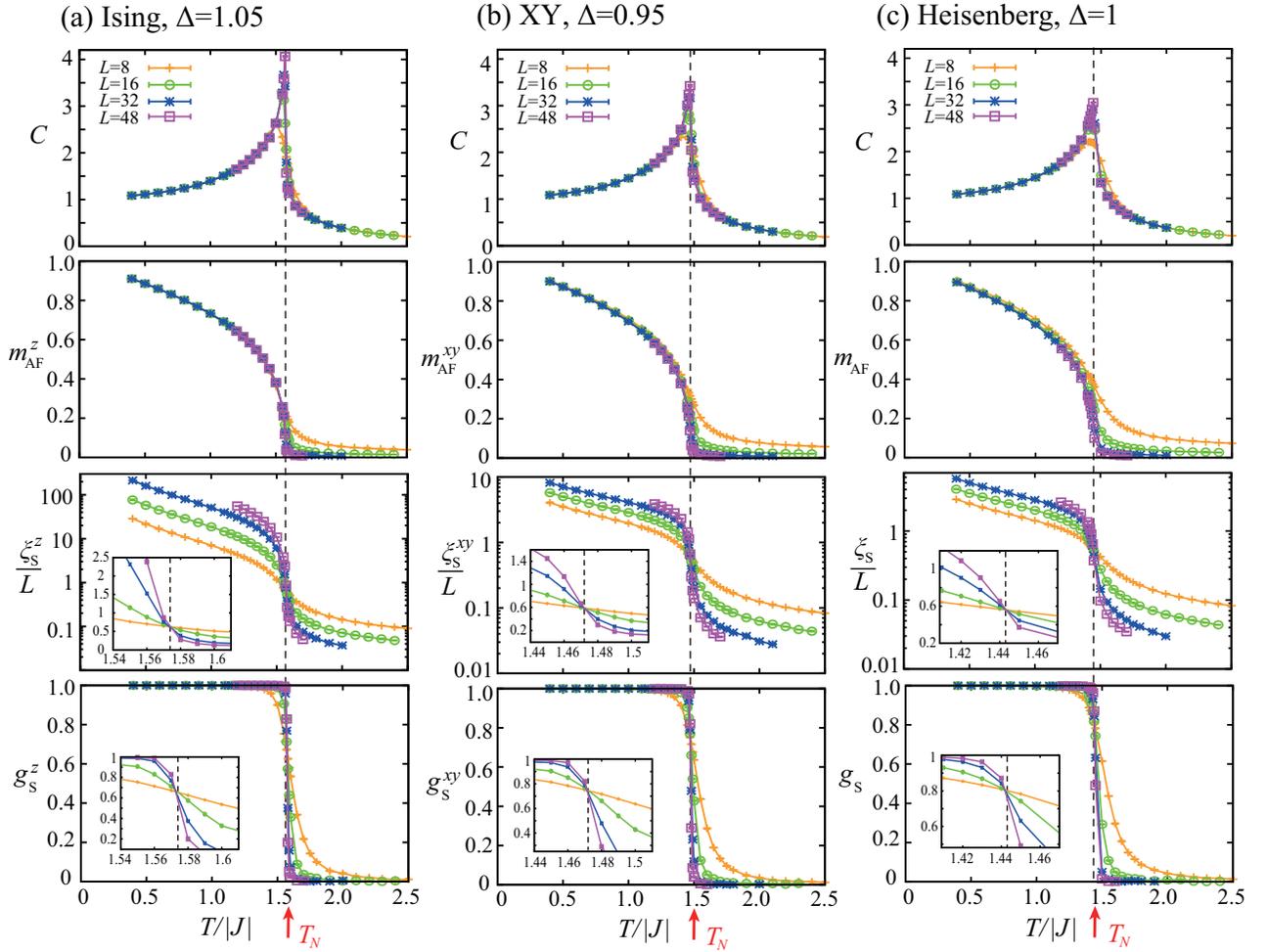}
\caption{MC results for the classical $XXZ$ model on the cubic lattice (\ref{eq:Hamiltonian}).  (a) Ising-type ($\Delta=1.05$), (b) $XY$-type ($\Delta=0.95$), and (c) Heisenberg-type ($\Delta=1$) spin systems. From top to bottom, the specific heat $C$, the antiferromagnetic order parameter , the ratio of the spin correlation length to the system size, and the Binder ratio are shown as a function of temperature. In lower two panels, an inset shows a zoomed view of each main panel near the magnetic transition temperature $T_N/|J|$ indicated by a dashed line. In (a), (b), and (c),  the transition temperatures are estimated as $T_N/|J|\simeq 1.574$, 1.472, and 1.443, respectively. \label{fig:basic_comp}}
\end{figure*}

Fundamental ordering properties of the classical antiferromagnetic $XXZ$ model on the cubic lattice (\ref{eq:Hamiltonian}) can be examined by means of the MC simulation. In our MC simulations, at each temperature, we perform $2\times 10^5$ MC sweeps and the first half is discarded for thermalization, where one MC sweep consists of the 1 heat-bath sweep and successive 10 over-relaxation sweeps. Observations are done in every MC sweep, and the statistical average is taken over 8 independent runs starting from different initial spin configurations. In the Ising-type, $XY$-type, and Heisenberg-type spin systems, the antiferromagnetic order parameters, the associated spin-correlation lengths, and the Binder ratios are respectively expressed as $m_{\rm AF}^z$, $\xi_s^z$, and $g_s^z$, $m_{\rm AF}^{xy}$, $\xi_s^{xy}$, and $g_s^{xy}$, and $m_{\rm AF}$, $\xi_s$, and $g_s$ which are defined by 
\begin{eqnarray}
m_{\rm AF}^z &=& \sqrt{ \langle G_2^z({\bf Q})\rangle }, \nonumber\\
m_{\rm AF}^{xy} &=& \sqrt{ \langle G_2^x({\bf Q}) + G_2^y({\bf Q})\rangle }, \nonumber\\
m_{\rm AF} &=& \sqrt{ \langle G_2^x({\bf Q}) + G_2^y({\bf Q}) + G_2^z({\bf Q}) \rangle } \nonumber\\
\xi_s^z &=& \frac{1}{2\sin(\pi/L)} \sqrt{\frac{\langle G_2^z({\bf Q})\rangle}{\langle G_2^z({\bf Q}+{\bf k}_{\rm min})\rangle}-1 }, \nonumber\\
\xi_s^{xy} &=& \frac{1}{2\sin(\pi/L)} \sqrt{\frac{\langle \sum_{\alpha=x,y}G_2^\alpha({\bf Q})\rangle }{\langle \sum_{\alpha=x,y}G_2^\alpha({\bf Q}+{\bf k}_{\rm min}\rangle)}-1 }, \nonumber\\
\xi_s &=& \frac{1}{2\sin(\pi/L)} \sqrt{\frac{\langle \sum_{\alpha=x,y,z}G_2^\alpha({\bf Q})\rangle }{\langle \sum_{ \alpha=x,y,z}G_2^\alpha({\bf Q}+{\bf k}_{\rm min})\rangle}-1 }, \nonumber\\
g_s^z &=& 2 \Big(1-\frac{1}{2}\frac{\langle G_2^z({\bf Q})^2\rangle}{ \langle G_2^z({\bf Q})\rangle^2} \Big), \nonumber\\
g_s^{xy} &=& \frac{3}{2} \Big(1-\frac{1}{3}\frac{\langle [G_2^x({\bf Q})+G_2^y({\bf Q})]^2\rangle}{\langle G_2^x({\bf Q})+G_2^y({\bf Q})\rangle^2} \Big), \nonumber\\
g_s &=& \frac{5}{2} \Big(1-\frac{3}{5}\frac{\langle [G_2^x({\bf Q})+G_2^y({\bf Q})+G_2^z({\bf Q})]^2\rangle}{\langle G_2^x({\bf Q})+G_2^y({\bf Q})+G_2^z({\bf Q})\rangle^2} \Big), \nonumber\\
G_2^\alpha({\bf q}) &=&  \Big|\frac{1}{N_{\rm spin}}\sum_i S^\alpha_i \, e^{i \, {\bf q}\cdot {\bf r}_i}\Big|^2 ,\nonumber\\
{\bf Q} &=& (\pi, \pi,\pi), \qquad {\bf k}_{\rm min} = (2\pi/L,0,0). \nonumber
\end{eqnarray}
Note that the ordering vector ${\bf Q} = (\pi, \pi,\pi)$ describes the two-sublattice antiferromagnetic order.

Figure \ref{fig:basic_comp} shows the temperature dependences of the specific heat $C$, the antiferromagnetic order parameter , the ratio of the spin correlation length to the system size $L$, and the Binder ratio for the Ising-type ($\Delta=1.05$), $XY$-type ($\Delta=0.95$), and Heisenberg-type ($\Delta=1$) spin systems. In all the three cases, the order parameters $m_{\rm AF}^z$, $m_{\rm AF}^{xy}$, $m_{\rm AF}$ start growing up at the antiferromagnetic transition temperatures $T_N$ indicated by specific-heat sharp peaks (see the upper two panels in Fig. \ref{fig:basic_comp}). The $T_N$ can accurately be determined as a cross point of different size data for the spin-correlation length ratio and Binder ratio. From the crossing points (see the lower two panels in Fig. \ref{fig:basic_comp}), we estimate $T_N$ as  $T_N/|J|\simeq 1.574$, 1.472, and 1.443 for $\Delta=1.05$, 0.95, and 1, respectively. Since exactly speaking, the crossing point depends on the choice of two different sizes, careful analysis of the system-size dependence of the crossing points is necessary to further accurately determine the transition temperature. Nevertheless, in the Heisenberg case of $\Delta=1$, the obtained value of $T_N/|J|=1.443$ is close to the so-far-reported best estimate of 1.457 \cite{3DHeisenberg_Campostrini_02}. In other two cases of $\Delta=1.05$ and 0.95, the corresponding estimates are not available.  

\section{Analytical calculations based on the linear spin-wave theory}
In the low-temperature long-range ordered phase, the magnetic excitations can be well described by the linear spin-wave theory (LSWT), so that we analytically investigate the temperature dependence of $\kappa_{\mu\nu}$ and $\sigma^s_{\mu \nu}$ based on the LSWT. Previously, we derived the corresponding result in the two-dimensional square-lattice system \cite{trans-sq_AK_prb_19}. Since the difference between two and three dimensions consists only in the dispersion relation and the momentum space integral, the formal expressions for various quantities before the ${\bf q}$-summation are basically the same as those in the two-dimensional system.

\subsection{Magnon representation}
The magnon representation of the Hamiltonian (\ref{eq:Hamiltonian}) and the spin and thermal currents in Eqs. (\ref{eq:j_sc}) and (\ref{eq:j_th}) can be derived by using the spin-wave expansions. In the Ising case of $\Delta > 1$, the quantization axis of spin is in the $S^z$ direction, and in the Heisenberg case of $\Delta=1$ where the quantization axis can be arbitrary, we chose it in the $S^z$ direction for simplicity. In the $XY$ case of $\Delta <1$ where the $S^x$ and $S^y$ components of spins are ordered, we take the quantization axis in the $S^x$ direction. To easily diagonalize the Hamiltonian (\ref{eq:Hamiltonian}) for $\Delta \geq 1$ ($\Delta <1$), we introduce the transformation from the laboratory frame to the rotated frame with $S^y$ ($S^z$) being the rotation axis, 
\begin{equation}
\left\{\begin{array}{l} 
S^{z (x)}_i= \tilde{S}^{z (x)}_i \cos(\theta_i) - \tilde{S}^{x (y)}_i \sin(\theta_i) \\
S^{x (y)}_i= \tilde{S}^{z (x)}_i \sin(\theta_i) + \tilde{S}^{x (y)}_i \cos(\theta_i) \\
S^{y (z)}_i= \tilde{S}^{y (z)}_i 
\end{array} \right . ,\nonumber
\end{equation} 
where $\theta_i = {\bf Q}\cdot {\bf r}_i$ and ${\bf Q}=(\pi,\pi,\pi)$ is the ordering vector of the two-sublattice antiferromagnetic order. By further using the Holstein-Primakoff transformation
\begin{equation}
\left\{\begin{array}{l} 
\tilde{S}^{z (x)}_i = S- \hat{a}^\dagger_i \hat{a}_i \\
\tilde{S}^{x (y)}_i + i \tilde{S}^{y (z)}_i = \sqrt{2S}\Big(1-\frac{\hat{a}^\dagger_i \hat{a}_i}{2S} \Big)^{\frac{1}{2}}\hat{a}_i = \sqrt{2S} \, \hat{a}_i +{\cal O}(S^{-\frac{1}{2}})\\
\tilde{S}^{x (y)}_i - i \tilde{S}^{y (z)}_i = \sqrt{2S}\hat{a}^\dagger_i\Big(1-\frac{\hat{a}^\dagger_i \hat{a}_i}{2S} \Big)^{\frac{1}{2}} = \sqrt{2S} \, \hat{a}^\dagger_i+{\cal O}(S^{-\frac{1}{2}}) \\
\end{array} \right . 
\end{equation}
with $\hat{a}^\dagger_i$ and $\hat{a}_i$ being respectively the bosonic creation and annihilation operators and the Fourier transformation of these operators 
\begin{equation}
\hat{a}^\dagger_i = \frac{1}{\sqrt{L^3}}\sum_{\bf q} \hat{a}^\dagger_{\bf q} e^{-i{\bf q}\cdot {\bf r}_i}, \quad \hat{a}_i = \frac{1}{\sqrt{L^3}}\sum_{\bf q} \hat{a}_{\bf q} e^{i{\bf q}\cdot {\bf r}_i},
\end{equation}
we obtain
\begin{eqnarray}
{\cal H} &=&\frac{1}{2}\sum_{\bf q}\Big[ A_{\bf q} \big( \hat{a}^\dagger_{\bf q}\hat{a}_{\bf q} + \hat{a}_{\bf q}\hat{a}^\dagger_{\bf q} \big)- B_{\bf q} \big(\hat{a}^\dagger_{\bf q}\hat{a}^\dagger_{-{\bf q}}+ \hat{a}_{\bf q}\hat{a}_{-{\bf q}} \big) \Big] \nonumber\\
&+& const. \nonumber
\end{eqnarray}
in the lowest order in the $1/S$ expansion. Here, the coefficients $A_{\bf q}$ and $B_{\bf q}$ are given by
\begin{eqnarray}\label{eq:AB}
A_{\bf q} &=& -6 J S \left\{\begin{array}{l} 
\Delta \qquad\qquad\qquad\quad (\Delta \geq 1) \\
1 - \frac{1}{2} (1-\Delta)\gamma_{{\bf q}} \quad (\Delta < 1) \\
\end{array} \right . , \nonumber\\
B_{\bf q} &=& -6 JS \left\{\begin{array}{l}
\gamma_{\bf q} \qquad\qquad\quad (\Delta \geq 1) \\
\frac{1}{2}(1+\Delta)\gamma_{{\bf q}} \quad (\Delta < 1) \\
\end{array} \right . , \nonumber\\
\gamma_{\bf q} &=& \frac{1}{3}\big[ \cos(q_x)+\cos(q_y)+\cos(q_z) \big] .
\end{eqnarray} 
The above Hamiltonian for the $\hat{a}_{\bf q}$ magnons can be diagonalized with the help of the Bogoliubov transformation
\begin{equation}
\left\{\begin{array}{l} 
\hat{a}_{\bf q} = U_{\bf q} \, \hat{b}_{\bf q} + V_{\bf q} \, \hat{b}^\dagger_{-{\bf q}}, \\
U_{\bf q}=U_{-{\bf q}} = \frac{1}{2}\Big[ \Big(\frac{A_{\bf q}+B_{\bf q}}{A_{\bf q}-B_{\bf q}} \Big)^{1/4} + \Big(\frac{A_{\bf q}-B_{\bf q}}{A_{\bf q}+B_{\bf q}} \Big)^{1/4}\Big], \nonumber\\
V_{\bf q}=V_{-{\bf q}} = \frac{1}{2}\Big[ \Big(\frac{A_{\bf q}+B_{\bf q}}{A_{\bf q}-B_{\bf q}} \Big)^{1/4} - \Big(\frac{A_{\bf q}-B_{\bf q}}{A_{\bf q}+B_{\bf q}} \Big)^{1/4}\Big],
\end{array} \right .  
\end{equation}
where $\hat{b}^\dagger_{\bf q}$ and $\hat{b}_{\bf q}$ are the creation and annihilation operators for magnons, and we obtain
\begin{equation}\label{eq:Hamiltonian_mag}
{\cal H} \simeq \sum_{\bf q} \varepsilon_{\bf q} \, \hat{b}^\dagger_{\bf q}\hat{b}_{\bf q}, \qquad \varepsilon_{\bf q} = \sqrt{A_{\bf q}^2-B_{\bf q}^2}, \nonumber\\
\end{equation} 
where we have dropped constant and higher-order terms.
In the $XY$ and Heisenberg cases of $\Delta\leq 1$, the magnon excitation is gapless, while in the Ising case of $\Delta > 1$, it has the excitation gap of $\Delta_{gp}=6|J|S\sqrt{\Delta^2-1}$, which can be seen from the following expression of $\varepsilon_{\bf q}$ near ${\bf q}=0$:
\begin{equation}\label{eq:magenergy_app}
\varepsilon_{\bf q} \simeq 6|J|S \left\{ \begin{array}{l}
\sqrt{\Delta^2-1 + \frac{1}{3} |{\bf q}|^2} \qquad (\Delta > 1) \\
\sqrt{\frac{1}{3}} \, |{\bf q}| \qquad\qquad\qquad \, (\Delta = 1) \\
\sqrt{ \frac{1}{6}(1+\Delta) } \, |{\bf q}| \qquad\quad \, (\Delta < 1) \\
\end{array} \right. . \nonumber 
\end{equation}
In the gapless cases of $\Delta \leq 1$, the magnon dispersion shows a ${\bf q}$-linear dependence, so that the magnon velocity ${\bf v}_{\bf q}=\nabla_{\bf q} \varepsilon_{\bf q}$ becomes $const. \times \hat{q}$ for the gapless mode.

In the same $1/S$ expansion, the thermal and spin currents in Eqs. (\ref{eq:j_th}) and (\ref{eq:j_sc}) can be expressed with the use of the $\hat{b}_{\bf q}$ magnons as follows:
\begin{equation}\label{eq:current_th_mag}
{\bf J}_{th} = \big( 6|J|S\big)^2 \sum_{\bf q} \tilde{\varepsilon}_{\bf q} \, \tilde{{\bf v}}_{\bf q}  \, \hat{b}_{\bf q}^\dagger \hat{b}_{\bf q} ,
\end{equation}
\begin{equation}\label{eq:current_spin_mag}
{\bf J}^z_s = \left\{\begin{array}{l} 
3 |J| S   \sum_{\bf q} \tilde{{\bf v}}_{\bf q}  \Big[  \frac{A_{\bf q}}{B_{\bf q}} ( \hat{b}_{\bf q}^\dagger \hat{b}_{-{\bf q}+{\bf Q}}^\dagger + \hat{b}_{\bf q} \hat{b}_{-{\bf q}+{\bf Q}} )\\
\qquad \qquad \qquad -( \hat{b}_{\bf q}^\dagger \hat{b}_{{\bf q}+{\bf Q}} + \hat{b}_{\bf q} \hat{b}_{{\bf q}+{\bf Q}}^\dagger )    \Big]    \quad (\Delta \geq 1) \\
{\cal O}\big( S^{1/2} \big) \qquad\qquad\qquad\qquad\qquad\qquad\quad (\Delta < 1) \\
\end{array} \right . 
\end{equation}
with 
\begin{equation}
\tilde{\varepsilon}_{\bf q} = \frac{\varepsilon_{\bf q}}{6|J|S}, \quad \tilde{{\bf v}}_{\bf q} =\nabla_{\bf q} \tilde{\varepsilon}_{\bf q} .
\end{equation}
In contrast to the thermal current ${\bf J}_{th}$ having the common magnon-representation, the spin current takes different forms depending on the value of $\Delta$. Of particular importance is that the spin currents in the $XY$ ($\Delta<1$) and other ($\Delta \geq 1$) cases are of the order of ${\cal O}\big( S^{1/2} \big)$ and ${\cal O}\big( S^{1} \big)$, respectively, which suggests that ${\bf J}^z_s$ in the $XY$ case is negligibly small as it is a higher order contribution in the $1/S$ expansion. Such a difference between $\Delta \geq 1$ and $\Delta<1$ cases stems from the fact that in the former and latter cases, the quantization axis of spin is parallel and perpendicular to the spin polarization of the spin current, respectively. Remember that although the spin current has its foundation on the conservation of the magnetization, only the $z$ component of the magnetization is conserved in the $XXZ$ model (\ref{eq:Hamiltonian}) with $\Delta \neq 1$. 

\subsection{Equal-time correlation function}
As the magnon Hamiltonian (\ref{eq:Hamiltonian_mag}) is already diagonalized, the partition function can easily be calculated as
\begin{equation}
Z = {\rm Tr} \Big[ \exp\big( - \frac{1}{T}\sum_{\bf q}\varepsilon_{\bf q} \hat{b}_{\bf q}^\dagger \hat{b}_{\bf q} \big) \Big] = \prod_{\bf q}\big[-f_{\rm B}(-\varepsilon_{\bf q}) \big] \nonumber
\end{equation}
with the Bose distribution function $f_{\rm B}(x)=(e^{x/T}-1)^{-1}$. Then, the equal-time correlation function $\langle J_{th,\nu}(0) \, J_{th,\mu}(0) \rangle$ for the thermal current whose magnon representation is given by Eq. (\ref{eq:current_th_mag}) can be calculated as
\begin{eqnarray}\label{eq:Jth_static_tmp}
&&\big\langle J_{th,\nu}(0) \, J_{th,\mu}(0) \big\rangle = \sum_{{\bf q},{\bf q}'} \varepsilon_{\bf q} \varepsilon_{{\bf q}'} v_{{\bf q},\mu} v_{{\bf q}',\nu} \big\langle \hat{b}_{\bf q}^\dagger \hat{b}_{\bf q} \hat{b}_{{\bf q}'}^\dagger \hat{b}_{{\bf q}'} \big\rangle \nonumber\\
&& \qquad = \delta_{\mu,\nu} \sum_{\bf q} \big[ \varepsilon_{{\bf q}} \, v_{{\bf q},\mu} \big]^2 f_{\rm B}(\varepsilon_{\bf q})\big[1+2f_{\rm B}(\varepsilon_{\bf q}) \big],
\end{eqnarray}
where we have used the formula $\big\langle \hat{b}_{\bf q}^\dagger \hat{b}_{\bf q} \hat{b}_{{\bf q}'}^\dagger \hat{b}_{{\bf q}'} \big\rangle = \frac{T^2}{Z} \frac{\partial ^2 \, Z}{\partial \varepsilon_{\bf q} \partial \varepsilon_{{\bf q}'}} $.

By taking the classical limit of
\begin{equation}\label{eq:classical_limit}
f_{\rm B}(x) \rightarrow \frac{T}{x},
\end{equation}
we obtain the equal-time correlation for the classical spins $\langle J_{th,\nu}(0) \, J_{th,\mu}(0) \rangle_{\rm cl}$ as
\begin{equation}\label{eq:Jth_static}
\big\langle J_{th,\nu}(0) \, J_{th,\mu}(0) \big\rangle_{\rm cl} = \delta_{\mu,\nu}  \, T^2  \, 2\sum_{\bf q} \big[  v_{{\bf q},\mu} \big]^2.
\end{equation}
At this point, the $T^{2}$ dependence of $\langle J_{th,\nu}(0) \, J_{th,\mu}(0) \rangle_{\rm cl}$ is clear. For completeness, we shall check whether $\sum_{\bf q} \big[  v_{{\bf q},\mu} \big]^2$ converges or not. 
Since in three dimensions, the ${\bf q}$-summation is written as
\begin{equation}\label{eq:qsum}
\sum_{\bf q} \simeq \frac{L^3}{(2\pi)^3} \int_0^{2\pi} d\phi_{\bf q}  \int_{-1}^{1} d\cos(\theta_{\bf q}) \int_{0}^{\pi}q^2 \, dq,
\end{equation} 
the problem is whether the $q$-integral of a physical quantity diverges or not in the $q \rightarrow 0$ limit. In the case of $\sum_{\bf q} \big[  v_{{\bf q},\mu} \big]^2$, $v_{{\bf q},\mu}$ does not diverge at ${\bf q}=0$ from the beginning, so that $\sum_{\bf q} \big[  v_{{\bf q},\mu} \big]^2$ converges, justifying the the $T^{2}$ dependence of $\langle J_{th,\nu}(0) \, J_{th,\mu}(0) \rangle_{\rm cl}$.

In the same manner, the temperature dependence of the equal-time correlation function for the spin current can be examined. Since in the $XY$ case of $\Delta<1$, the spin current is absent within the leading-order magnon contribution [see Eq. (\ref{eq:current_spin_mag})], we only consider the $\Delta \geq 1$ case in which after some manipulations, we have
\begin{eqnarray}\label{eq:Js_static_pretmp}
&&\big\langle J^z_{s,\nu}(0) \, J^z_{s,\mu}(0) \big\rangle = \frac{-1}{4}\sum_{{\bf q},{\bf q}'} v_{{\bf q},\nu} v_{{\bf q}',\mu} \Big\{\big( \delta_{{\bf q},{\bf q}'} + \delta_{{\bf q},{\bf q}'+{\bf Q}}\big) \nonumber\\
&& \quad \times \big[ f_{\rm B}(\varepsilon_{\bf q})f_{\rm B}(-\varepsilon_{{\bf q}+{\bf Q}})+f_{\rm B}(-\varepsilon_{\bf q})f_{\rm B}(\varepsilon_{{\bf q}+{\bf Q}}) \big]\nonumber\\
&& \, -\frac{A_{\bf q}}{B_{\bf q}}\frac{A_{{\bf q}'}}{B_{{\bf q}'}}\big( \delta_{{\bf q},{\bf q}'} + \delta_{{\bf q},-{\bf q}'+{\bf Q}}\big) \nonumber\\
&&\quad \times \big[f_{\rm B}(\varepsilon_{\bf q})f_{\rm B}(\varepsilon_{-{\bf q}+{\bf Q}})+f_{\rm B}(-\varepsilon_{\bf q})f_{\rm B}(-\varepsilon_{-{\bf q}+{\bf Q}}) \big] \Big\}.
\end{eqnarray}
Now, we take the classical limit of Eq. (\ref{eq:Js_static_pretmp}). As the relations, $A_{\pm{\bf q}+{\bf Q}}=A_{\bf q}$, $B_{\pm{\bf q}+{\bf Q}}=-B_{\bf q}$, and ${\bf v}_{\pm{\bf q}+{\bf Q}}=\pm {\bf v}_{\bf q}$, are satisfied for $\Delta\geq 1$, the classical limit Eq. (\ref{eq:classical_limit}) yields
\begin{equation}\label{eq:Js_static_tmp}
\big\langle J^z_{s,\nu}(0) \, J^z_{s,\mu}(0) \big\rangle_{\rm cl}=\delta_{\mu,\nu} \, T^2 \, \sum_{\bf q} \big[ v_{{\bf q},\mu}\big]^2 \Big( 1+\frac{A_{\bf q}^2}{B_{\bf q}^2}\Big)\frac{1}{\varepsilon_{\bf q}^2}.
\end{equation}
In the ${\bf q}\rightarrow 0$ limit, $B_{\bf q}\rightarrow -6JS$ and $1/\varepsilon^2_{\bf q}\rightarrow 1/(\Delta^2-1+\frac{1}{3}q^2)$ [see Eqs. (\ref{eq:AB}) and (\ref{eq:magenergy_app})], so that even for $\Delta=1$, $\sum_{\bf q} \big[ v_{{\bf q},\mu}\big]^2 \Big( 1+\frac{A_{\bf q}^2}{B_{\bf q}^2}\Big)\frac{1}{\varepsilon_{\bf q}^2} \propto \int_0^\pi q^2 \frac{1}{q^2}dq$ is non-divergent. Thus, for $\Delta \geq 1$, $\big\langle J^z_{s,\nu}(0) \, J^z_{s,\mu}(0) \big\rangle_{\rm cl}$ exhibits the $T^2$ dependence similarly to $\big\langle J_{th,\nu}(0) \, J_{th,\mu}(0) \big\rangle_{\rm cl}$. 

\subsection{Spin and thermal conductivities}  
In the classical spin systems, the conductivities $\kappa_{\mu\nu}$ and $\sigma^s_{\mu\nu}$ are obtained from the time-correlation of the associated currents [see Eq. (\ref{eq:conductivity})]. To calculate the time correlation, it is convenient to start from the quantum mechanical system and take the classical limit of Eq. (\ref{eq:classical_limit}) afterwards. In the quantum mechanical system, the dynamical correlation function $L^{a,b}_{\mu\nu}(\omega)$ can be expressed in the following form \cite{book_AGD}:
\begin{eqnarray}
L^{a,a}_{\mu\nu}(\omega) &=& -\frac{Q^{a,R}_{\mu\nu}(\omega)-Q^{a,R}_{\mu\nu}(0)}{i\omega}, \nonumber\\
Q^{a,R}_{\mu\nu}(\omega) &=& Q^a_{\mu\nu}(\omega + i 0), \\ 
Q^a_{\mu\nu}(i\omega_n) &=& -\frac{1}{L^3}\int_0^{1/T}\big\langle T_{\tau} J_{a,\mu}(\tau)J_{a,\nu}(0) \big\rangle \, e^{i \omega_n \, \tau} d\tau. \nonumber
\end{eqnarray} 
Here, $Q^a_{\mu\nu}(i\omega_n)$ is a response function and $\omega_n=2\pi n T$ is the bosonic Matsubara frequency. Then, the thermal conductivity $\kappa_{\mu\nu}$ and the spin conductivity $\sigma^s_{\mu\nu}$ are given by
\begin{eqnarray}\label{eq:conductivity_quantum}
\kappa_{\mu\nu} &=&\frac{1}{T} i \frac{d \, Q^{th,R}_{\mu\nu}(\omega)}{d \, \omega} \Big|_{\omega =0}, \nonumber\\
\sigma^s_{\mu\nu} &=& i \frac{d \, Q^{s,R}_{\mu\nu}(\omega)}{d \, \omega} \Big|_{\omega =0}.
\end{eqnarray}

We first calculate the magnon thermal conductivity $\kappa_{\mu\nu}$ for which the response function $Q^{th}_{\mu\nu}(i\omega_n)$ is given by \cite{book_AGD}
\begin{eqnarray}
Q^{th}_{\mu\nu}(i\omega_n) &=& \frac{-1}{L^3}\sum_{\bf q} \varepsilon_{\bf q}^2 v_{{\bf q},\mu} \, v_{{\bf q},\nu} \, T\sum_{\omega_m}{\cal D}_{\bf q}(i\omega_m){\cal D}_{\bf q}(i\omega_m+i\omega_n) \nonumber\\
&=& \frac{-1}{L^3}\sum_{\bf q} \varepsilon_{\bf q}^2 v_{{\bf q},\mu} \,  v_{{\bf q},\nu} \int_{-\infty}^{\infty}\frac{dx}{2\pi i} \big[ {\cal D}^R_{\bf q}(x)-{\cal D}^A_{\bf q}(x) \big] \nonumber\\
&& \times  \big[ {\cal D}^R_{\bf q}(x+i\omega_n) + {\cal D}^A_{\bf q}(x-i\omega_n) \big] \, f_{\rm B}(x) ,
\end{eqnarray}
where ${\cal D}^R_{\bf q}(x)$ (${\cal D}^A_{\bf q}(x)=\big[{\cal D}^R_{\bf q}(x)\big]^*$) is the retarded (advanced) magnon Green's function obtained by analytic continuation $i\omega_m \rightarrow \omega + i0$ in the temperature Green's function ${\cal D}_{\bf q}(i\omega_m)$ defined by 
\begin{equation}
{\cal D}_{\bf q}(\tau) = -\big\langle T_{\tau} \hat{b}_{\bf q}(\tau)\hat{b}_{\bf q}^\dagger(0) \big\rangle = T\sum_{\omega_m} {\cal D}_{\bf q}(i\omega_m) \, e^{-i\omega_m \tau}.
\end{equation} 
With the use of Eq. (\ref{eq:conductivity_quantum}), the thermal conductivity in the quantum system is formally expressed as
\begin{equation}\label{eq:conductivity_quantum_th}
\kappa_{\mu\nu}=\frac{T^{-1}}{4\pi L^3}\int_{-\infty}^\infty dx \sum_{\bf q} \varepsilon_{\bf q}^2 \, v_{{\bf q},\mu} \, v_{{\bf q},\nu} f_{\rm B}'(x)\big[ {\cal D}^R_{\bf q}(x)-{\cal D}^A_{\bf q}(x) \big]^2.
\end{equation}
Here, the magnon Green's function ${\cal D}_{\bf q}^R(x)$ is given by
\begin{equation}\label{eq:Green_mag}
{\cal D}_{\bf q}^R(x) = \frac{1}{x-\varepsilon_{\bf q}+i \alpha \, x} = \big[ {\cal D}_{\bf q}^A(x) \big]^\ast,
\end{equation}
where the dimensionless coefficient $\alpha$ represents the magnon damping \cite{MagnonTrans_Tatara_15}. In the present system where the Hamiltonian (\ref{eq:Hamiltonian}) involves only the spin variable, the damping $\alpha$ is brought by the magnon-magnon scatterings. 
The temperature dependence of $\alpha$ will be discussed below.

In the classical spin system with $f_{\rm B}'(x)=-T/x^2$ [see Eq. (\ref{eq:classical_limit})], by substituting Eq. (\ref{eq:Green_mag}) into Eq. (\ref{eq:conductivity_quantum_th}), we obtain the following expression for the thermal conductivity in the classical spin systems $\kappa_{\mu \nu}^{\rm cl}$ as
\begin{equation}
\kappa_{\mu \nu}^{\rm cl} = \frac{1}{2L^3}\frac{1+\alpha^2}{\alpha}\sum_{\bf q}\frac{1}{\varepsilon_{\bf q}} \, v_{{\bf q},\mu} \, v_{{\bf q},\nu},
\end{equation}
where the equation
\begin{equation}\label{eq:integral}
\int_{-\infty}^\infty \frac{dx}{ \big[(x-\varepsilon_{\bf q})^2+(\alpha x)^2\big]^2 } = \frac{\pi}{2}\frac{1+\alpha^2}{\varepsilon_{\bf q}^3 \alpha^3}
\end{equation}
has been used. As the ${\bf q}$-summation $\sum_{\bf q}\frac{1}{\varepsilon_{\bf q}} \, v_{{\bf q},\mu} \, v_{{\bf q},\nu}$ turns out to converge even in the gapless cases of $\Delta \leq 1$ where the summation is proportional to $\delta_{\mu,\nu} \int_0^\pi q^2 \frac{1}{q} dq$, $\kappa_{\mu \nu}^{\rm cl}$ can be expressed as
\begin{equation}\label{eq:conductivity_classical_th}
\kappa_{\mu \nu}^{\rm cl} \simeq \delta_{\mu,\nu}\frac{1+\alpha^2}{\alpha} \times const. 
\end{equation}
irrespective of the value of $\Delta$. In the lower temperature region where the magnon damping is sufficiently small such that $\alpha \ll 1$, it follows that $\kappa_{\mu\nu}^{\rm cl}\propto 1/\alpha$, which agrees with the results obtained in other theoretical approaches \cite{MagnonTrans_Tatara_15, MagnonTrans_Jiang_13}.
Thus, in all the three ($\Delta > 1$, $\Delta=1$, and $\Delta<1$) cases, the temperature dependence of $\kappa_{\mu\mu}^{\rm cl} \propto 1/\alpha$ is governed by the magnon damping factor $\alpha$.

The damping of the antiferromagnetic magnon due to multi-magnon scatterings has already been calculated by using Feynman diagram techniques in Refs. \cite{MagnonDamping_Harris_71}. The temperature dependence of $\alpha$ in the classical Heisenberg antiferromagnet essentially follows the $T^2$ form, i.e., $\alpha \propto T^2$, which results from the leading-order scattering process involving four magnons. In the $XY$-type and Ising-type classical spin systems, although the concrete expression of $\alpha$ is not available, the same temperature dependence $\alpha \propto T^2$ is expected because the same types of the Feynman diagrams (the same leading-order scattering processes) contribute to the magnon damping. Of course, there must be quantitative differences among the three cases. In particular, for the Ising-type anisotropy of $\Delta>1$, the magnon excitation is gapped, so that the phase space satisfying the energy conservation in the calculation of the relevant Feynman diagrams would be shrunk with increasing $\Delta$, resulting in a smaller value of $\alpha$. Apart from such a quantitative difference which may become serious for strong Ising-type anisotropies, the longitudinal thermal conductivity $\kappa_{\mu\mu}^{\rm cl}$ in the classical limit should behave as $\kappa_{\mu\mu}^{\rm cl} \propto 1/\alpha \propto 1/T^2$ in all the three ($\Delta > 1$, $\Delta=1$, and $\Delta<1$) cases.       

We next calculate the spin conductivity $\sigma^s_{\mu\nu}$ based on Eq. (\ref{eq:conductivity_quantum}). As in the case of $\kappa_{\mu\nu}$, starting from the magnon representation of the spin current in Eq. (\ref{eq:current_spin_mag}), we can write down the response function $Q^s_{\mu\nu}(i\omega_n)$ as
\begin{eqnarray}
&& Q^s_{\mu\nu}(i\omega_n) = \frac{-1}{4L^3}\sum_{{\bf q},{\bf q}'} v_{{\bf q},\mu} \, v_{{\bf q}',\nu} \, \Big\{ \big(\delta_{{\bf q},{\bf q}'}+\delta_{{\bf q},{\bf q}'+{\bf Q}} \big) F^+_{\bf q}(i \omega_n) \nonumber\\
&&\qquad\qquad\quad + \frac{A_{\bf q}}{B_{\bf q}}\frac{A_{{\bf q}'}}{B_{{\bf q}'}}\big(\delta_{{\bf q},{\bf q}'}+\delta_{{\bf q},-{\bf q}'+{\bf Q}} \big)F^-_{\bf q}(i \omega_n) \Big\} , \nonumber\\
&& F^\pm_{\bf q}(i \omega_n) = T\sum_{\omega_m} {\cal D}_{\bf q}(i\omega_m) \big[ {\cal D}_{{\bf Q}\pm {\bf q}}(i\omega_n\pm i\omega_m) \nonumber\\
&& \qquad\qquad\qquad\qquad\qquad\qquad + {\cal D}_{{\bf Q}\pm{\bf q}}(-i\omega_n\pm i\omega_m) \big] \nonumber\\
&& = \int_{-\infty}^\infty \frac{dx}{2\pi i} f_{\rm B}(x)\Big\{ \big[ {\cal D}^R_{{\bf Q}\pm {\bf q}}(\pm x+i \omega_n) + {\cal D}^A_{{\bf Q}\pm {\bf q}}(\pm x-i \omega_n) \big] \nonumber\\
&& \times \big[ {\cal D}^R_{\bf q}(x)- {\cal D}^A_{\bf q}(x) \big]  \pm \big[ {\cal D}^R_{{\bf Q}\pm {\bf q}}(\pm x) - {\cal D}^A_{{\bf Q}\pm {\bf q}}(\pm x) \big] \nonumber\\
&& \times \big[ {\cal D}^R_{\bf q}(x+i\omega_n) + {\cal D}^A_{\bf q}(x-i\omega_n) \big]  \Big\}. 
\end{eqnarray}
Then, the spin conductivity $\sigma^s_{\mu \nu}$ is formally written as
\begin{eqnarray}\label{eq:conductivity_quantum_spin}
&&\sigma^s_{\mu\nu} = \frac{1}{8\pi L^3 }\int_{-\infty}^\infty dx \sum_{{\bf q},{\bf q}'} v_{{\bf q},\mu} \, v_{{\bf q}',\nu} \, f_{\rm B}'(x) \big[ {\cal D}^R_{\bf q}(x)- {\cal D}^A_{\bf q}(x) \big] \nonumber\\
&&\times \Big\{ \big(\delta_{{\bf q},{\bf q}'}+\delta_{{\bf q},{\bf q}'+{\bf Q}} \big)  \big[ {\cal D}^R_{{\bf q}+{\bf Q}}(x)- {\cal D}^A_{{\bf q}+{\bf Q}}(x) \big] \\
&&-\frac{A_{\bf q}}{B_{\bf q}}\frac{A_{{\bf q}'}}{B_{{\bf q}'}} \big(\delta_{{\bf q},{\bf q}'}+\delta_{{\bf q},-{\bf q}'+{\bf Q}} \big) \big[ {\cal D}^R_{-{\bf q}+{\bf Q}}(x)- {\cal D}^A_{-{\bf q}+{\bf Q}}(x) \big] \Big\}. \nonumber
\end{eqnarray}
In the same manner as that for $\kappa_{\mu\nu}$, we will derive the spin conductivity in the classical limit $\sigma^{s,{\rm cl}}_{\mu \nu}$.
By substituting Eq. (\ref{eq:Green_mag}) into Eq. (\ref{eq:conductivity_quantum_spin}), taking the classical limit of $f_{\rm B}'(x)=-T/x^2$, and using Eq. (\ref{eq:integral}) and the formula
\begin{equation}
\int_{-\infty}^\infty \frac{dx}{ \big[(x-\varepsilon_{\bf q})^2+(\alpha x)^2\big]\big[(x+\varepsilon_{\bf q})^2+(\alpha x)^2\big] } = \frac{\pi}{2}\frac{1}{\varepsilon_{\bf q}^3 \alpha}, \nonumber
\end{equation}
we have 
\begin{equation}
\sigma^{s,{\rm cl}}_{\mu\nu} = \frac{1}{2L^3}T\sum_{\bf q} v_{{\bf q},\mu} \, v_{{\bf q},\nu} \frac{1}{\varepsilon_{\bf q}^3}\Big[ \frac{1+\alpha^2}{\alpha}+ \alpha \frac{A_{\bf q}^2}{B_{\bf q}^2} \Big]. 
\end{equation}
In the Ising case of $\Delta>1$,  the ${\bf q}$-summation $\sum_{\bf q} v_{{\bf q},\mu} \, v_{{\bf q},\nu} \frac{1}{\varepsilon_{\bf q}^3}$ converges, while not in the Heisenberg case of $\Delta=1$ because the ${\bf q}$-summation involves $\int_0^\pi q^2 \frac{1}{q^3} dq$ which yields the logarithmic divergence. Thus, we could summarize the result as follows:
\begin{equation}\label{eq:conductivity_classical_spin}
\sigma^{s,{\rm cl}}_{\mu\nu} \simeq \delta_{\mu,\nu} \left\{ \begin{array}{l}
\displaystyle{T\Big[ c_1 \alpha + c_2 \, \frac{1}{\alpha}  \Big]   \quad (\Delta > 1)} \\
\displaystyle{T c_3 \int_0^\pi \frac{1}{q} dq  \qquad (\Delta =1)}   \nonumber\\
\displaystyle{0  \qquad\qquad \qquad  \, (\Delta <1) }  \nonumber\\
\end{array} \right. 
\end{equation}
with constants $c_1$, $c_2$, and $c_3$.
In contrast to the thermal conductivity $\kappa_{\mu\nu}^{\rm cl}$, the spin conductivity $\sigma^{s,{\rm cl}}_{\mu \nu}$ reflects the difference in the ordering properties. First of all, in the $XY$ case of $\Delta<1$, $\sigma^{s, {\rm cl}}_{\mu\nu}$ is zero because the spin current is absent within the leading-order magnon contribution [see Eq. (\ref{eq:current_spin_mag})]. In the Ising case of $\Delta>1$, the temperature dependence of $\sigma^{s, {\rm cl}}_{\mu\mu}$ is determined by that of $T/\alpha$ at sufficiently low temperatures where $\alpha \ll1$ is expected. Since for relatively weak anisotropies, $\alpha \propto T^2$ is expected to be satisfied, the longitudinal spin conductivity should exhibit the following temperature dependence: $\sigma^{s,{\rm cl}}_{\mu\mu} \propto T/\alpha \propto T^{-1}$. In the Heisenberg case of $\Delta=1$, due to the logarithmic divergence, the longitudinal spin conductivity should remain infinite over the low-temperature ordered phase where the LSWT is applicable.  


\end{document}